\documentclass[11pt]{article}

\usepackage{jheppub}
\usepackage[utf8]{inputenc}
\usepackage{booktabs}
\usepackage{graphicx}
\usepackage{subcaption}
\usepackage{mathtools}
\usepackage{cancel}
\usepackage{datetime}
\usepackage[dvipsnames]{xcolor}
\usepackage{slashed}
\usepackage{tcolorbox}
\usepackage{multirow}

\usepackage{soul} 

\numberwithin{equation}{section}

\def\ash{%
  {
    a%
    \kern-.045em%
    \lower.3ex\hbox{sh}%
}%
}

\newcommand{\nn}{\nonumber}

\usepackage{xcolor}

\makeatletter
\newcommand{\Biggg}{\bBigg@{3.5}}
\makeatother

\newcommand{\be}{\begin{equation}}
\newcommand{\ee}{\end{equation}}
\newcommand{\bea}{\begin{eqnarray}}
\newcommand{\eea}{\end{eqnarray}}

\begin{document}

\title{Higgsless simulations of cosmological phase transitions 
	and gravitational waves}
\author{Ryusuke Jinno$^a$,}
\author{Thomas Konstandin$^b$,}
\author{Henrique Rubira$^c$,}
\author{Isak Stomberg$^b$}

\affiliation{\small$^a$ Instituto de F\'{\i}sica Te\'orica IFT-UAM/CSIC, C/ Nicol\'as Cabrera 13-15, Campus de Cantoblanco, 28049, Madrid, Spain}
\affiliation{\small$^b$ Deutsches Elektronen-Synchrotron DESY, Notkestr.~85, 22607 Hamburg, Germany}
\affiliation{\small$^c$ Physik Department T31, Technische Universit\"at M\"unchen,
	James-Franck-Stra\ss e 1, D-85748 Garching, Germany}

\preprint{DESY 22-148, IFT-UAM/CSIC-22-100, TUM-HEP-1416/22}

\abstract{First-order cosmological phase transitions in the early Universe source sound waves and, subsequently, a background of stochastic gravitational waves. Currently, predictions 
	of these gravitational waves rely heavily on simulations of a Higgs field coupled to the plasma of the early Universe, the former providing the latent heat of the phase transition.
	Numerically, this is a rather demanding task since several length scales enter the dynamics. 	
From smallest to largest, these are the thickness of the Higgs
	interface separating the different phases, the shell thickness of the sound waves, and
	the average bubble size. In this work, we present an approach to perform Higgsless simulations in three dimensions, producing fully nonlinear results, while at the same time removing the hierarchically smallest scale from the lattice.
This significantly reduces the complexity of the problem and contributes to making our approach highly efficient.
	We provide spectra for the produced gravitational waves for various choices of wall velocity and strength of the phase transition, as well as introduce a fitting function for the spectral shape.
}

\maketitle

\newpage
\section{Introduction}
\label{sec:Introduction}

After the direct detection of gravitational waves (GWs) from compact binaries by LIGO-Virgo~\cite{Abbott:2016blz,Abbott:2016nmj,TheLIGOScientific:2017qsa}, the next milestone experiment in GW science will be LISA \cite{Seoane:2013qna, amaroseoane2017laser,LISACosmologyWorkingGroup:2022jok}. While LIGO-Virgo was designed to probe GWs in the kHz frequency band sourced by binaries with a few tens of solar masses, LISA will focus on GWs from supermassive binaries in the mHz frequency band. Besides detecting local sources such as black holes and neutron stars, LISA will offer the chance to probe a stochastic GW background (SGWB) emitted in the early Universe. This is exciting in light of the conclusion in~\cite{Grojean:2006bp}, namely that a first-order phase-transition (PT) occurring when the temperature of the Universe was around the electroweak scale (TeV) could source mHz GWs, providing an appealing scenario to be probed by LISA.

A first-order electroweak PT is motivated by different extensions of the standard model of particle physics. In the case in which the PT is of the first-order kind, the scalar field does not transition smoothly between the phases. It means that thermal or quantum fluctuations in the Higgs sector could induce different points of space to locally tunnel from the metastable vacuum to the global vacuum \cite{Coleman:1977py, Linde:1980tt, Steinhardt:1981ct}. Those regions of true vacuum expand symmetrically in all directions forming bubbles that eventually collide sourcing GWs \cite{Witten:1984rs}. Though the collision of the scalar-field solitons is known as an important contribution to the GW spectra \cite{Kosowsky:1991ua,Kosowsky:1992rz,Kosowsky:1992vn,Kamionkowski:1993fg,Huber:2008hg,Konstandin:2017sat}, it was shown in \cite{Hindmarsh:2013xza} that the case in which the Higgs sector transfers energy to the surrounding plasma when the bubbles expand also characterizes an important GW source. In that case, the expansion of the bubbles pushes the surrounding plasma forming sound waves that propagate for a long time. The overlap of many of those sound shells induces a UV peak in the GW spectra that grows linearly in time\footnote{The typical lifetime for the propagation of sound waves is determined by the onset of turbulence or Hubble damping~\cite{Caprini:2009yp,Hindmarsh:2013xza,Hindmarsh:2015qta,Hindmarsh:2017gnf, Hindmarsh:2016lnk,Hindmarsh:2019phv,Ellis:2019oqb,Auclair:2022jod}. Turbulence also constitutes an important source of GWs.}.

In spite of analytical models providing useful insights about the GWs from sound-waves \cite{Caprini:2009fx,Hindmarsh:2016lnk,Jinno:2016vai,Jinno:2017fby, Hindmarsh:2019phv}, numerical simulations are vital to capturing nonlinearities (e.g.~shocks) in the primordial fluid. The current state-of-the-art for simulations embraces two different approaches. First, the Helsinki-Sussex group, which solves the hydro+scalar field system in the lattice, coupling both sectors via a phenomenological friction \cite{Hindmarsh:2013xza,Hindmarsh:2015qta,Hindmarsh:2017gnf,Cutting:2019zws}. Second, the hybrid (and already Higgsless) approach presented in \cite{Jinno:2020eqg}, by part of the authors of this article, in which we embedded a 1D spherically-symmetric hydro simulation into a 3D lattice. The Higgs field in that case was included as a space- and time-dependent boundary condition in the 1D simulation using the bag equation of state \cite{Espinosa:2010hh,Giese:2020rtr,Giese:2020znk}. Since the bubbles have cosmological radius (of the order of the Hubble radius) and the dynamics of the Higgs field occurs on TeV scale, there are at least 30 orders of magnitude between those two physics scales. For all effects, the Higgs field is local in space and time for the bubbles. Moreover, since typically one can only span over three orders of magnitude in a lattice, simulating the Higgs together with the bubbles introduces an additional source of systematic errors.

In this work, we present a full 3D Higgsless simulation to calculate the sound shell contribution to the GW spectra. We profit from having a hydro simulation in which the expanding bubbles are treated as a non-dynamical background and the shock waves in the plasma are resolved by an appropriate numerical scheme~\cite{KURGANOV2000241}. Again, the Higgs field couples to the plasma via conservation of the stress-energy tensor, locally changing the enthalpy and energy balance as in \cite{Jinno:2020eqg}. The fluid dynamics is solved fully nonlinearly on a 3D lattice. 

The structure of this work is as follows. We start in Sec.~\ref{sec:basic} by presenting the basic setup for the fluid evolution and the calculation of the GW spectrum. Then in Sec.~\ref{sec:num}, we describe a high-resolution numerical scheme used to solve the fluid equations. In Sec.~\ref{sec:res} we present the numerical results for the GW spectrum. Sec.~\ref{sec:disc} is devoted to discussion and conclusions.

\section{Basic setup}
\label{sec:basic}

In this section, we discuss the basic setup. We start by discussing the construction of a bubble nucleation history. We then move on to explain the fluid dynamics and how to incorporate the latent heat that is released into the plasma, without the need to keep track of the Higgs dynamics itself. Finally, we discuss GWs produced by sound waves. 

\subsection{Bubble nucleation histories}

The main assumption in our approach is that the expansion velocity of the Higgs bubbles is 
given as an external parameter. This allows simulating bubble histories (nucleation times and locations)
that are consistent with the assumption of an exponentially increasing bubble nucleation probability
\be
P \simeq P_0 \exp(\beta (t-t_0)) \, .
\ee

We briefly describe our method to generate these histories~\cite{Jinno:2020eqg}. First, consider the case 
that the bubbles expand with the speed of light, $v_w=1$. 
The easiest way to produce such histories is to homogeneously distribute the nucleation of $N_\tau$ bubbles in 
a four-volume consisting of the three space dimensions and the `time' $\tau = \exp[\beta (t-t_0)]$.
Nucleations that happen in the forward light-cone of other nucleations are removed and only a smaller 
number $N_b$ are truly part of the nucleation history. For larger values of $\tau$, all bubbles will nucleate in the forward light-cone of other bubbles and the phase transition has finished and $N_b$ will not increase anymore. 

It is worthwhile to notice that the outcome of this algorithm does not depend on the size of the volume in $\tau$ (which we always choose $\tau \in [0,1]$) or the number $N_\tau$ of bubbles that are homogeneously and randomly distributed as long as $N_\tau$ is large enough. 
Consider what happens when $N_\tau$ is doubled. The first bubble in average nucleates at $\tau\sim1/N_\tau$ and due to the doubling the phase transition starts earlier by $\Delta t \simeq \ln(2)/\beta$. All bubbles with $\tau < 1/2$ are also homogeneously distributed, and if $N_\tau/2$ was already large enough, the phase transition has already ended at $\tau = 1/2$ and percolation is finished. Thus, all bubbles with $\tau > 1/2$ lie in the forward light cone of other bubbles and are rejected. 

In essence, as long as $N_\tau \gg N_b$, an increase in $N_\tau$ only shifts the nucleations to slightly smaller times (likewise a change in the $\tau$ interval is inconsequential). The number of bubbles that pass the light-cone test  scales as
\be
N_b \simeq \frac{1}{8 \pi} \left( \frac{L\, \beta}{v_w} \right)^3 \, ,
\ee
where $L$ is the box size of the simulation and $v_w$ is the wall velocity. The bubble count $N_\tau$ should then be chosen significantly larger than that\footnote{The number of bubbles as well as their spatial distribution (and therefore the GW spectra) can be affected by temperature fluctuations during their nucleation. See \cite{Jinno:2021ury} for a description of how the GW spectra can be enhanced by up to a couple of orders of magnitude in that case.}.

Finally, notice that if distances are measured in terms of the bubble wall velocity $v_w$, 
consistent bubble nucleation histories can be deduced from the corresponding ones with $v_w=1$.
The method above works unchanged if all the bubble locations and the box size $L$ are rescaled with $v_w$.

\subsection{Fluid equations of motion} \label{sec:fluid}

We aspire to solve the equations corresponding 
to the energy-momentum conservation in the fluid
\be \label{eq:Tcons}
\partial_\mu T^{\mu\nu} = 0 \, ,  
\ee
with 
\begin{equation} \label{eq:Tmunu}
T^{\mu\nu} = u^\mu u^\nu w - g^{\mu\nu} p \, ,
\end{equation}
where $u^\mu = \gamma (1, \vec v)$ denotes the fluid four-velocity ($\gamma = 1/\sqrt{1-v^2}$), $w$ is the enthalpy and $p$ the pressure in the system.
Besides the constant bubble expansion velocity, the second assumption that enters our simulations 
concerns the equation of state. For simplicity, we abide by the bag equation of state~\cite{Espinosa:2010hh}, given as
\be
p = \frac13 a T^4 - \epsilon, \, \quad w = T\frac{dp}{dT}= \frac43 a T^4 \, .
\ee
The bag constant $\epsilon$ will be different in the symmetric and broken phases.
Hence it is a spacetime-dependent function that depends on the 
nucleation locations and times of the individual bubbles, $\epsilon(\vec x, t)$. 
We comment below on how to incorporate more complex equations of state in our framework.

In principle, the dynamics of the Higgs field could be quite complicated, since friction 
forces and local temperature changes have an impact on the expansion velocity of the boundary 
between the two phases. As mentioned before, we consider only the simplified case where the bubble walls 
expand with a fixed wall velocity $v_w$ that can be changed as an external parameter. 
Moreover, the numbers of degrees of freedom might be different in the two phases, leading to different 
constants $a$ in the equation of state. However, we will see that this will not enter in our 
framework (but it would be e.g.~important for entropy considerations).

The Euler equation is a conservation law. The four conserved quantities are
\be\label{K_def}
K^\mu := T^{\mu0} \, ,
\ee
such that 
\begin{equation} 
K^\mu = wu^0u^\mu -p\, \delta^{0\mu}\,,
\end{equation}
and the differential equations from Eq.~(\ref{eq:Tcons}) read
\begin{eqnarray}
\partial_t K^0 + \nabla_i K^i &=& 0 \, , \label{eq:K0cons} \\
\partial_t K^j + \nabla_i T^{ij}[K^\mu] &=& 0 \, , \label{eq:Kicons}
\end{eqnarray}
which then requires expressing the spatial components $T^{ij}$ of the energy-momentum 
tensor in terms of $K^\mu$. From Eq.~(\ref{eq:Tmunu}), we have
\be
K^0 = \gamma^2 w - \frac14 w + \epsilon \, ,
\ee
and
\be
\sum_i (K^i)^2 = \gamma^4 v^2 w^2 \, .
\ee
Inverting these quadratic relations one finds
\be
w = \frac43 \bar K^0 (2 \sqrt{1-\lambda} - 1) \, ,\quad 
v^2 = \frac{6 - 3 \lambda - 6\sqrt{1-\lambda}}{\lambda} \, ,
\ee
where we introduced 
\bea
\bar K^0 := K^0 - \epsilon \, \quad \text{and} \quad \lambda := \frac{3}{4 (\bar K^0)^2} \sum_i (K^i)^2 \, .
\eea
A slight problem with this expression is that in the limit of $K^i \to 0$ also $\lambda \to 0$, which makes this expression numerically unstable. This is easily remedied by the equivalence
\be
v^2 = \frac{6 - 3 \lambda - 6\sqrt{1-\lambda}}{\lambda} = \frac{3 \lambda}{(1 + \sqrt{1 - \lambda})^2}.
\ee
In the simulation, it is handy to express $T^{ij}$ directly in terms of $K^i$ 
\be
\label{eq:AnsatzTij}
T^{ij} = K^i K^j F + \delta^{ij} p \, ,
\ee
and one finds that the constant $F$ is given by 
\be
F = \frac{3}{2\bar K^0} \frac{1}{1 +  \sqrt{1 - \lambda}} \, ,
\ee
and again
\bea
w &=& \frac43 \bar K^0 (2 \sqrt{1 - \lambda} - 1) \, , \\
p &=& \frac{w}4 - \epsilon \, , \\
v^2 &=& \frac{3 \lambda}{(1 + \sqrt{1 - \lambda})^2} \, . 
\eea
Notice that the Ansatz (\ref{eq:AnsatzTij}) uses $K^i \parallel u^i \parallel v^i$ while in general $K^\mu \nparallel u^\mu$. Hence $T^{\mu\nu}$ can not be parametrized using $K^\mu$ in an analogous way.

The phase transition is then triggered by the fact that the bag constant $\epsilon(x,t)$
depends on space and time and changes for a grid point
in the simulation when the first bubble sweeps over this location. Our framework relies on performing 
the calculation of $T^{ij}(K^\mu, \epsilon)$ 
efficiently in the case of the bag equation of state. More involved equations of state
would have to modify this part, probably at the cost of solving the corresponding 
equations numerically or using an interpolation of $F(K^\mu, \epsilon)$.

\subsection{Gravitational wave spectrum}
\label{sec:GWs}

We follow the approach in Ref.~\cite{Jinno:2020eqg} to calculate the GW spectrum. In particular we ultimately measure the (dimensionless) GW power 
\be
Q'(q) = \frac{q^3 \beta}{w^2 V T} \int \frac{d\Omega_k}{4 \pi} 
\left[ \Lambda_{ij,kl} T_{ij} (q,\vec k) T_{kl}^* (q,\vec k) \right]_{q=k} \, ,
\ee
where $w$ denotes the enthalpy (before the phase transition), $q$ is the frequency, and $\vec k$ the momentum of the 
GW waves, $\Lambda$ is the projection on the TT part of the energy-momentum tensor $T_{ij}$, $V$ the volume, and $T$
the simulation time. 

This relates to the observed GW power spectrum according to~\cite{Jinno:2020eqg} 
\be
\frac{\Omega_{GW}}{Q'} = \frac{4H \tau_{\rm sw}}{3\pi^2} \frac{H}{\beta} \, ,
\label{eq:OmQp}
\ee
where $\tau_{\rm sw}$ is the duration of the sound waves that is either limited by decay into turbulence or Hubble damping, depending 
on the properties of the sound waves.

Notice that there is a subtlety in the projection when done in Fourier space~\cite{Jinno:2020eqg}. 
Due to the cyclicity on the grid, Fourier modes with momenta $N-1$ and $-1$ are equivalent. 
The lattice momentum $\vec l$ can be mapped to physical momentum $\vec k$ via
\be
\label{eq:mom1}
\vec k = 2 \frac{N}{L} \sin\left(\frac{\pi \vec l}{N}\right) \, ,
\ee
or
\be
\label{eq:mom2}
\vec k = \frac{N}{L} \sin\left(\frac{2\pi \vec l}{N}\right) \, ,
\ee
and which one should be used depends on context. For the choice (\ref{eq:mom1}) the lattice modes with norm $N-1$ and $1$ are associated with the 
same physical momentum, which only makes sense when the sign of the momentum is irrelevant. For the choice (\ref{eq:mom2}),
modes with norm $N/2$ and $0$ are associated with the same physical momentum, which only makes sense when the UV contribution 
is suppressed. 

For example, when the contribution to the isotropic GW spectrum from the various momentum modes is calculated, only the absolute value of the momentum is relevant, $q=|\vec k|$. Furthermore one would like to avoid that modes with $l \sim N/2$ contaminate the IR part of the spectrum (like in the fermion doubling problem). Hence, (\ref{eq:mom1}) is the preferred choice. 

On the other hand, in the projection operator $\Lambda_{ij, kl}$, the signs of the momenta are essential to obtain the 
correct projection for modes with $l \sim N$. Hence, (\ref{eq:mom2}) is the preferred choice. Still, the projection for the modes with $l \sim N/2$ will not be appropriate and we will have to discard the UV tail of the spectrum.

\section{Numerical method}
\label{sec:num}

Since we do not keep track of the Higgs field and essentially send the thickness of the Higgs wall to zero, we expect 
discontinuities in the fluid variables at the phase boundaries\footnote{Once more, we do not keep track of the Higgs dynamics, which is not a limitation but reduces the model-dependence of our framework. We discuss this further in Sec.~\ref{sec:disc}.}. The basis for our numerical setup is the method by Kurganov and Tadmor~\cite{KURGANOV2000241} (in the following KT) to solve conservation laws using a numerical central scheme with high resolution. In this subsection, we briefly put this method into context and explain its main building blocks.  

\subsection{The Kurganov-Tadmor scheme} \label{sec:KTscheme}

Solving conservation laws -- or partial differential equations (PDE) in general -- numerically is a vivid 
field of research. The complexity comes from the fact that many of these systems show advective 
behavior and allow for shocks and 
other features like rarefaction or compression waves. Often, this hinders the application 
of simple differencing schemes due to a balance 
between (excessive) numerical viscosity and unphysical oscillatory solutions. 

For example, the first-order (in time) Lax-Friedrich scheme leads to large numerical viscosity that 
will wash out any sharp shock front. However, constructing a second-order (in time) scheme 
(e.g.~the Lax-Wendroff scheme~\cite{Lax-Wendroff}) typically leads to local extrema in the solution and eventually to unphysical negative values
for the pressure and/or energy densities. This issue is highlighted by Godunov's theorem which states 
that this trade-off is unavoidable in linear numerical schemes. 

One way out of this conundrum is to solve the locally linearized problem exactly.
The functions that describe the fluid are 
assumed to be piecewise constant or piecewise linear with discontinuities at the cell borders. 
Evolving these functions exactly then requires to locally diagonalize a matrix in order to identify the left and right moving eigenvectors of the system. This is essentially the 
approach of Godunov schemes~\cite{godunov_scheme} or Riemann solvers. As expected, this spectral analysis is quite time-consuming 
even if done iteratively or even approximately.

An alternative route is to artificially introduce nonlinear terms -- even if the original 
PDE was linear. This process is called hybridization, and is most simply implemented using slope or flux limiters. 
For example, if the conservation law is expressed in terms of fluxes, one can limit the fluxes in 
a way to fulfill the so-called {\em total variation diminishing (TVD)} criterion. This criterion 
ensures that no local extrema are generated at the cost of some additional computations of the limiters. 
The limiters introduce a nonlinearity into the scheme, hence avoiding Godunov's theorem.

The main motivation of Kurganov and Tadmor was to find a Riemann-solver-free scheme using limiters.
One of the facets 
of Godunov's theorem is that the Lax Friedrich scheme cannot be easily promoted to second-order accuracy 
(in time). This is because the additional viscous terms in the equations diverge in the limit $\Delta t \to 0$.
Accordingly, KT designed a discretization scheme based on piecewise linear functions 
that allows for a semi-discrete formulation (i.e.~the limit $\Delta t \to 0$ is finite).
The only additional information needed is hereby the maximal eigenvalue $\rho$ of the linearized system. 
In our case,
\be
\rho \left( \frac{\partial T^{i\nu}}{\partial K^\sigma} \right) 
:=
{\rm max} \left\{ \left| \Lambda \left( \frac{\partial T^{i\nu}}{\partial K^\sigma} \right) \right| \right\} \, ,
\ee
where $\Lambda$ is the set of eigenvalues of the matrix $(...)^\nu_\sigma$.
The index $i$ is here an external index depending on which direction the flux is considered in.

In the following, we will use the notation of KT to describe the algorithm. Consider a one-dimensional PDE of the form
\be
\label{eq:PDE}
\partial_t u + \partial_x f(u) = 0 \, . 
\ee
The discretized version of this PDE according to KT is then
\be
u_j^{n+1} = u_j^n - \frac{\Delta t}{\Delta x} (H^n_{j+1/2} - H^n_{j-1/2}) \, ,
\ee
where $j$ denotes the index of the spatial grid and $n$ the index of the time grid. The flux function $H$ is evaluated in between the cells, and is given by (we dropped the time index $n$ since KT is an explicit scheme)
\be
\label{eq:flux}
H_{j+1/2} = \frac{f(u^+_{j+1/2})+f(u^-_{j+1/2})}{2} - \frac{a_{j+1/2}}{2} \left[ u^+_{j+1/2}- u^-_{j+1/2} \right] \, ,
\ee
where the staggered values of the fluid are given by
\be
u^+_{j+1/2} = u_{j+1} - \frac{\Delta x}{2} (u_x)_{j+1} \,, \quad
u^-_{j+1/2} = u_{j} + \frac{\Delta x}{2} (u_x)_{j} \, .
\ee
The velocities are determined using the minmod limiter
\be
(u_x)_j = {\rm minmod} 
\left(
\theta \frac{u_j - u_{j-1}}{\Delta x},
\frac{u_{j+1} - u_{j-1}}{{ 2} \Delta x},
\theta \frac{u_{j+1} - u_{j}}{\Delta x}
\right)
\ee
with a free parameter $\theta$ bound by $\theta \in [1,2]$ and $a$ denotes the maximal local velocity
\be
a_{j+1/2} = 
{\rm max} \left\{
\rho\left(\frac{\partial f}{\partial u}(u^+_{j+1/2})\right),
\rho\left(\frac{\partial f}{\partial u}(u^-_{j+1/2})\right)
\right\} \, .
\ee
The minmod limiter selects the element with the smallest absolute value if all elements have the same sign and is 
zero otherwise. 

We see that this scheme allows for a semi-discrete representation since the limit $\Delta t \to 0$ is finite. Hence, 
second-order accuracy (in time) can be achieved by using Runge-Kutta integration. In our simulations, 
we use the standard second-order relations that, for an equation of the form
\be
\frac{du}{dt} = g(u) = -\partial_x f\, ,
\ee
is given by
\be
u^{n+1} = u^n + \frac{\Delta t}6 (k_1 + 2 k_2 + 2 k_3 + k_4)\,,
\ee
with 
\bea
\label{eq:k}
k_1 &=& g(t^n, \, u^n) \, , \nn  \\
k_2 &=& g(t^n + \Delta t/2, \, u^n + \Delta t \, k_1/2) \, , \nn \\
k_3 &=& g(t^n + \Delta t/2, \, u^n + \Delta t \, k_2/2) \, , \nn \\
k_4 &=& g(t^n + \Delta t, \, u^n + \Delta t \, k_3) \, .
\eea
Those expressions (\ref{eq:PDE})--(\ref{eq:k}) generalize to several dimensions with three spatial indices $(j,k,l)$ and staggered indices $(j\pm 1/2,k,l)$, $(j,k\pm 1/2,l)$ and $(j,k,l\pm 1/2)$ for the three different fluxes~\cite{KURGANOV2000241}.

The second term in the flux (\ref{eq:flux}) acts as a viscous term. At the same time, the limiter fulfills the
TVD criterion and ensures local monotonicity of the solutions. The numerical viscosity is much smaller than in the 
case of Lax-Friedrich due to the dependence on the velocities $a_{j+1/2}$, leading to less dissipation and allowing our system to keep track of the shock solutions (see Sec.~\ref{sec:single_bubble}). These velocities are typically 
anyway calculated to test the Courant--Friedrichs--Lewy stability condition and hence no additional costs are incurred.

\vskip 0.5cm

In our setup using the bag equation of state, the local velocities can be determined analytically using the inversion relations (\ref{eq:AnsatzTij}). 
For the flux in the $x$-direction, two eigenvalues have the value $3 v_x/2/(1+\kappa)$ while the other two have somewhat lengthy closed expressions. 
In the limit of small fluid velocities, the two largest eigenvalues become $\pm c_s = \pm \sqrt{1/3}$, independent of the fact whether a discontinuity is close by or not.  
This allows working with a fixed step size in time according to the 
Courant--Friedrichs--Lewy stability condition. In the following, we also use this 
leading order expression and simply set $a_{j+1/2} = c_s$ in the KT scheme. We tested in several setups that 
choosing the full expression did not change our results, but stronger phase transitions might require special care.

\subsection{Single-bubble simulations} \label{sec:single_bubble}

We now move on to study the evolution of a single bubble, which is useful for various reasons. First, to show that by setting the single-bubble boundary conditions $\epsilon(\vec x,t)$ we can reproduce the self-similar solution for the $w$ and $v$ profiles seen in \cite{Espinosa:2010hh}. Reproducing this self-similar solution indicates the robustness of our approach. Second, to test the accuracy of our simulations and to study to which extent the shocks are resolved in our lattice.

\begin{figure}
	\centering
	\includegraphics[width=0.45\textwidth]{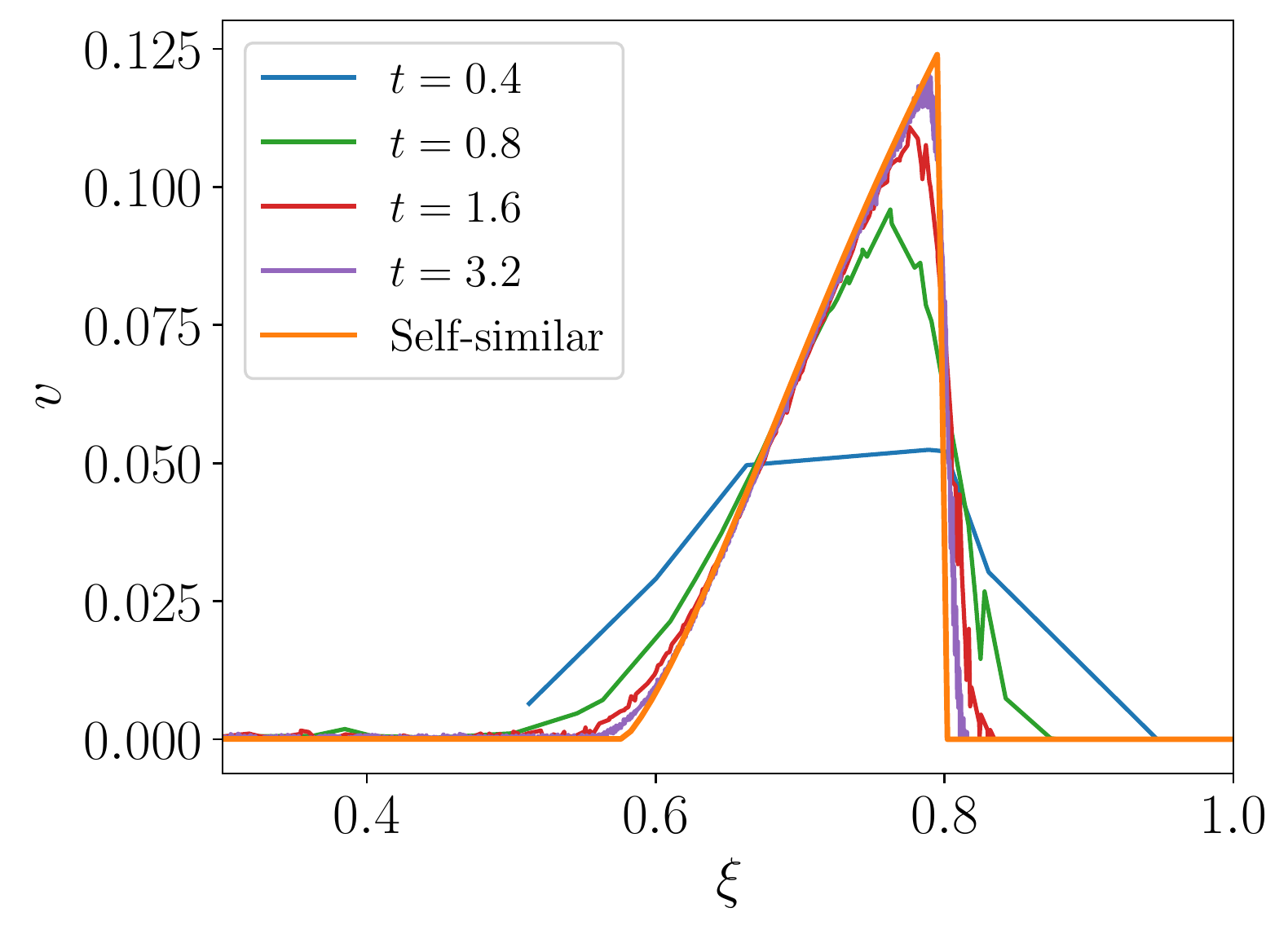} 
	\includegraphics[width=0.45\textwidth]{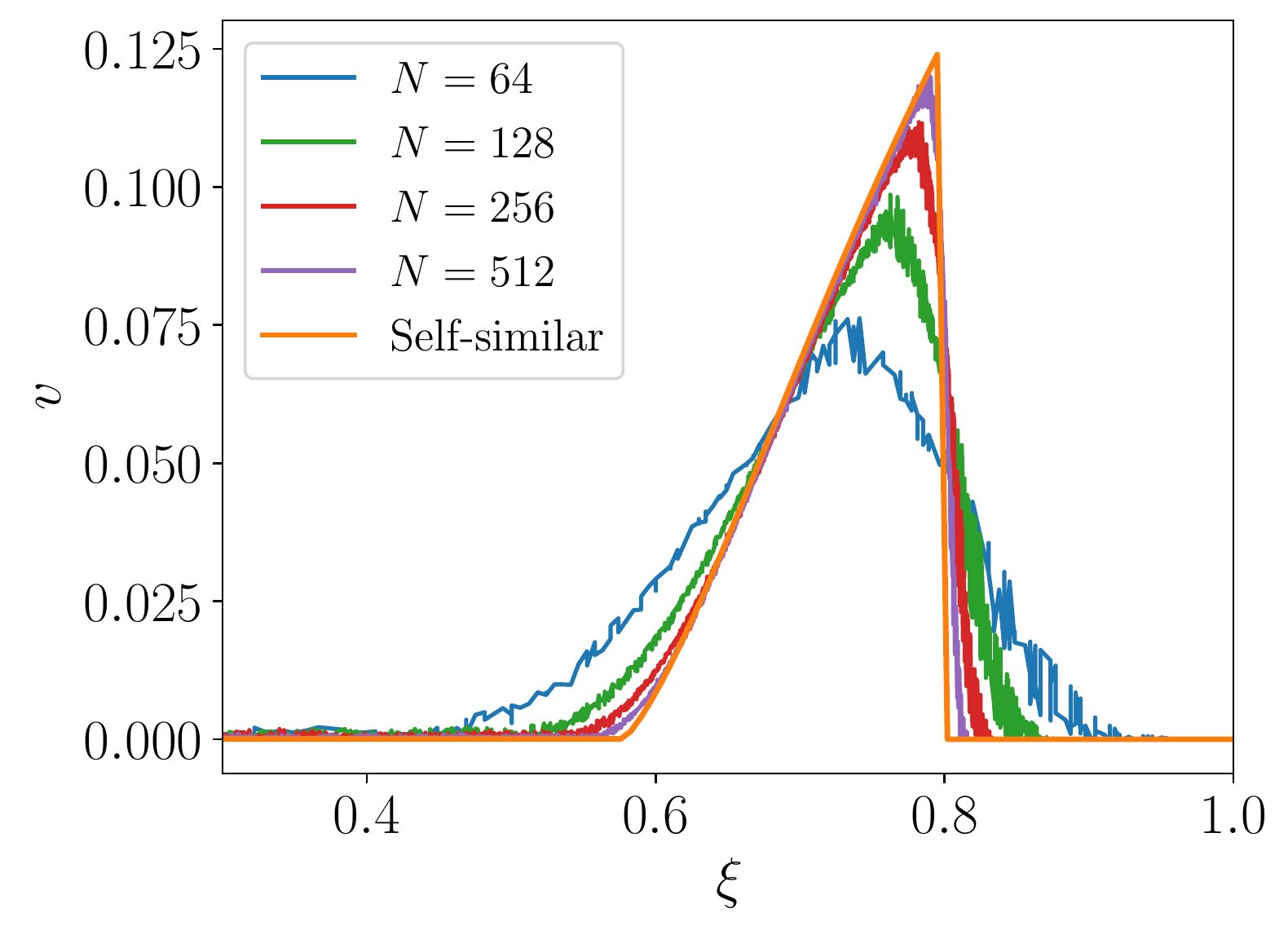} 
	\caption{On the left, the time evolution of the velocity profile is plotted. We nucleate a single bubble in a box with size $L=20v_w/\beta$, grid size $N = 512$, wall velocity $v_w = 0.8$ and transition strength $\alpha = 0.05$. On the right, the velocity profile is given as a function of $\xi$ for various values of $N$ at $t = 3.2/\beta$ for the same values of $L$, $v_w$ and $\alpha$. In both panels, we choose $100,000$ random points from the lattice and rescale them by the respective evolution time to obtain $\xi$. The self-similar solution of \cite{Espinosa:2010hh} is indicated in orange.
	}
	\label{fig:self_evolve}
\end{figure}

In the left panel of Fig.~\ref{fig:self_evolve} we display the time evolution of the velocity profile for a detonation in a grid with $N=512$ and $L=20v_w/\beta$. We take $100,000$ random points from the lattice and rescale them by the evolution time to get $\xi := r/t$. In the first time steps we do not see a shock front: there is a transient stage in which the lattice reacts to the time evolution of the initial conditions $\epsilon(\vec x,t)$. As time evolves, shocks are formed and maintained due to the robustness of the KT scheme described in Sec.~\ref{sec:KTscheme}.
Notice that for values of $N$ and $L$ used to calculate the GW spectrum in Sec.~\ref{sec:res}, the shocks have already formed at $t \gtrsim 1.6/\beta$. The first nucleated bubbles (those that contribute more to the GW spectrum) typically evolve for times longer than that value before colliding, indicating that shocks have already developed at average collision time and that the colliding wall profiles are already self-similar.

The right panel of Fig.~\ref{fig:self_evolve} displays the velocity profile at $t=3.2/\beta$ for different values of $N$. We see that the transient stage before developing the shocks takes longer for smaller $N$. The similarity between the left and right panels of Fig.~\ref{fig:self_evolve} is consistent with what is expected from the self-similar evolution: either doubling time evolution or doubling the grid resolution leads to the same effect. 

\begin{figure}
	\centering
	\includegraphics[width=0.45\textwidth]{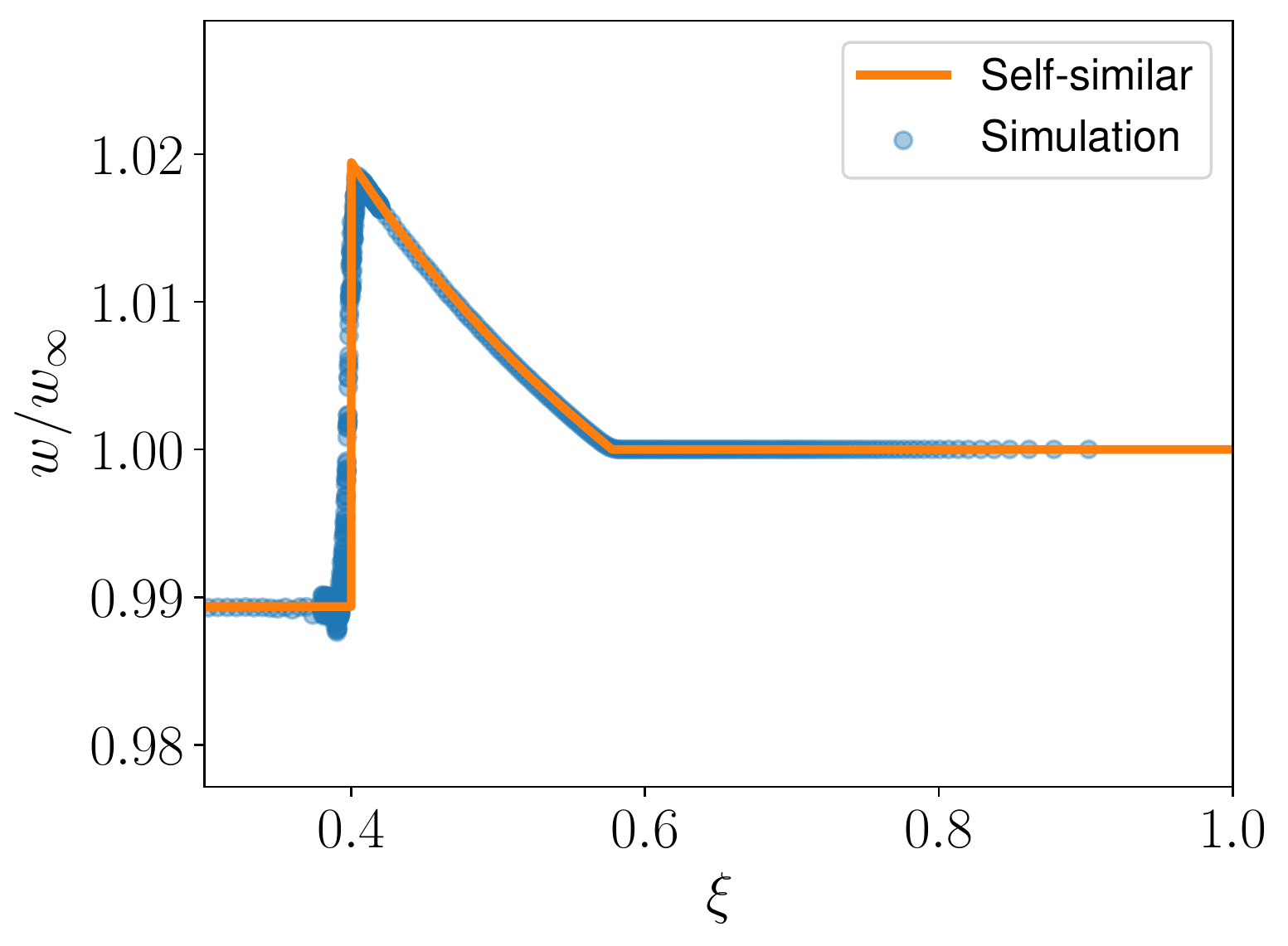} 
	\includegraphics[width=0.45\textwidth]{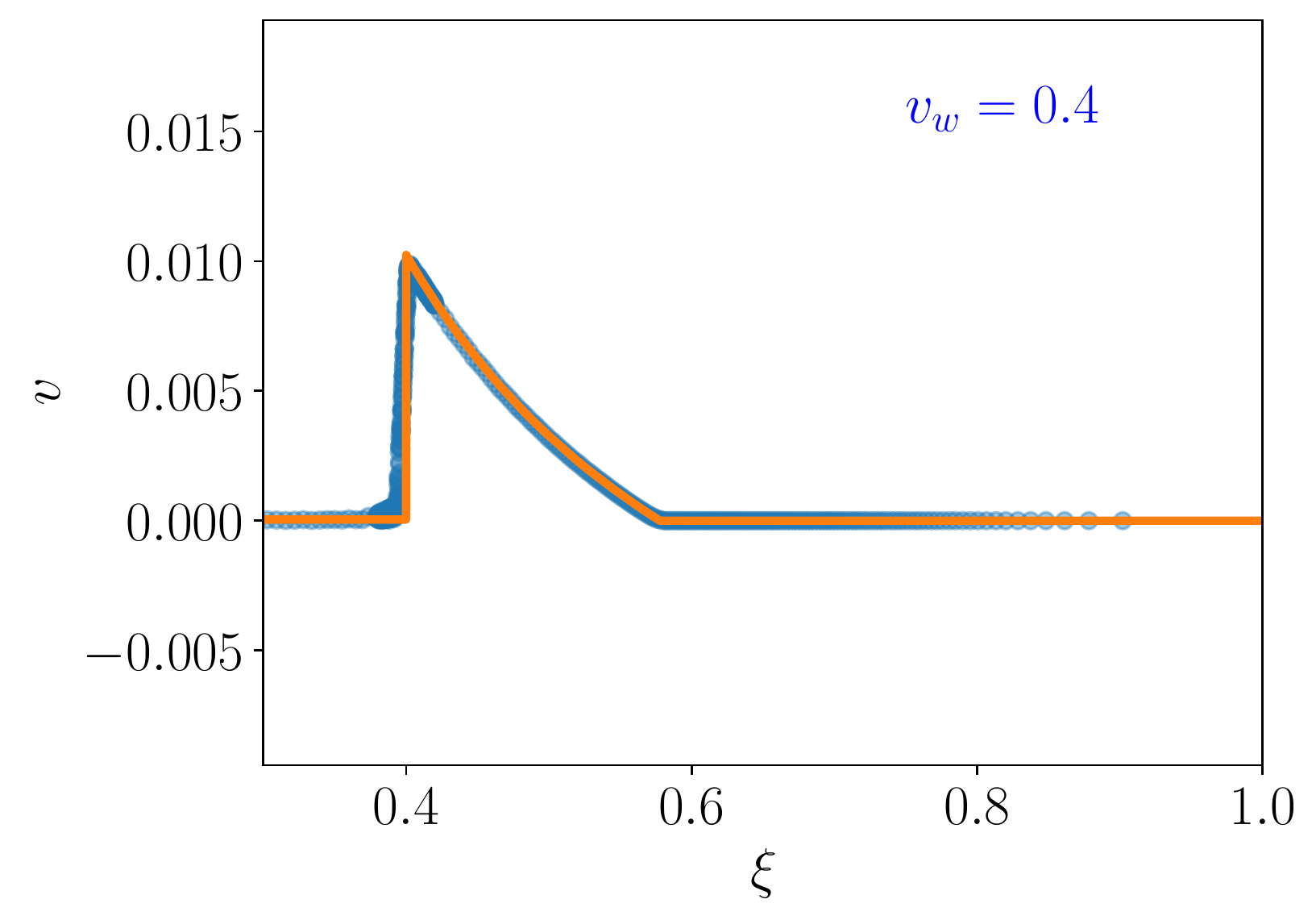} 
	\includegraphics[width=0.45\textwidth]{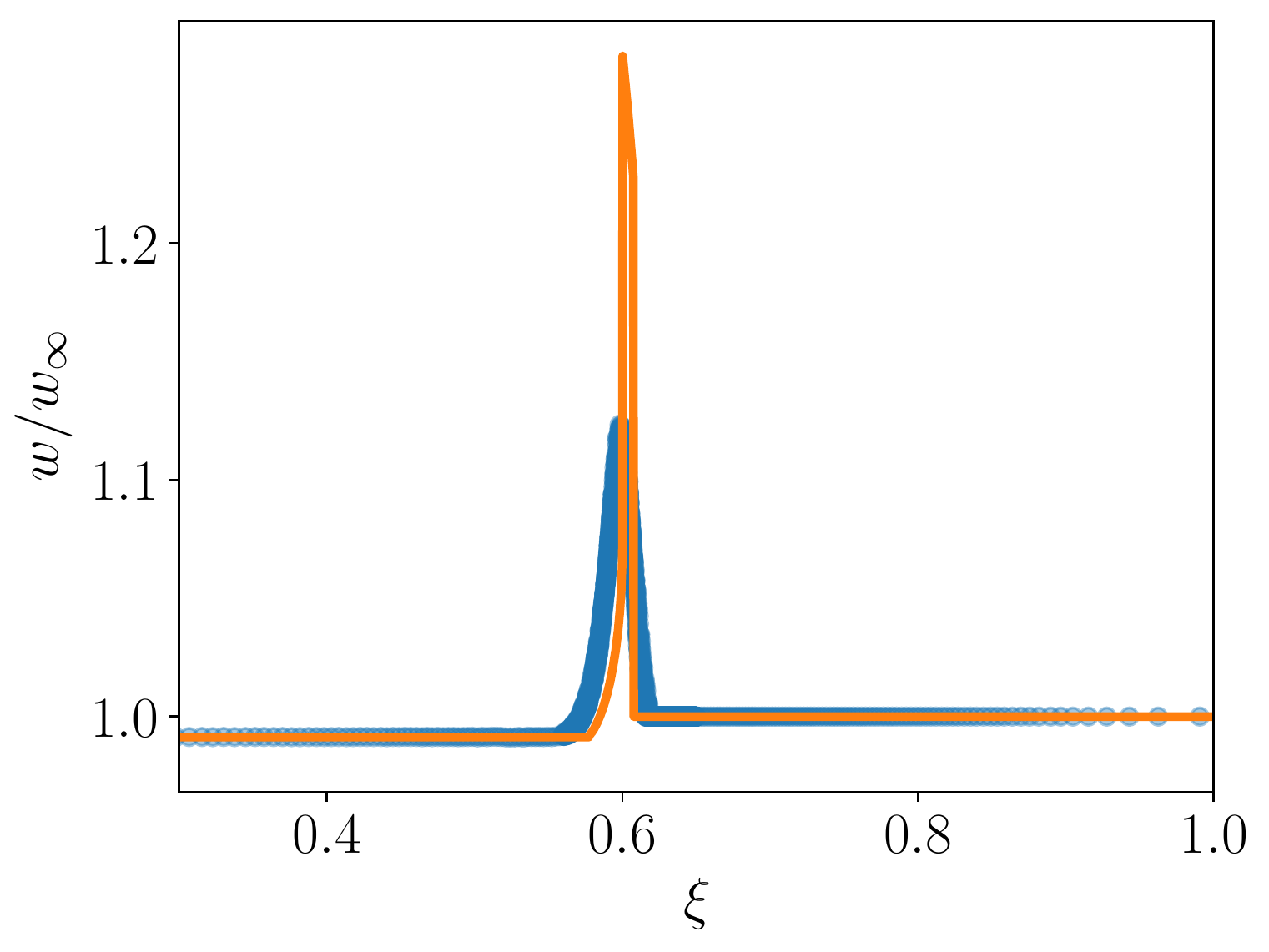} 
	\includegraphics[width=0.45\textwidth]{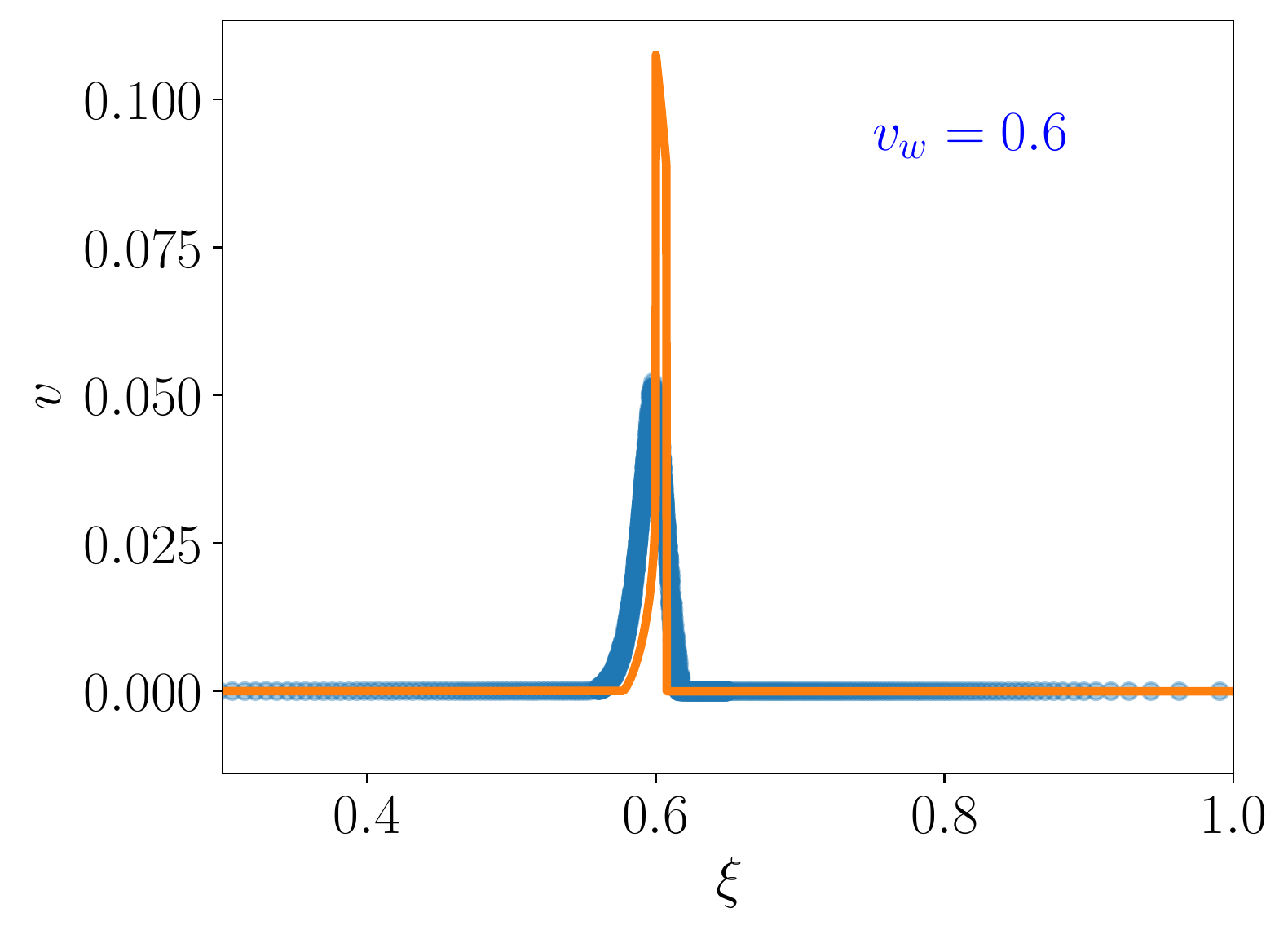} 
	\includegraphics[width=0.45\textwidth]{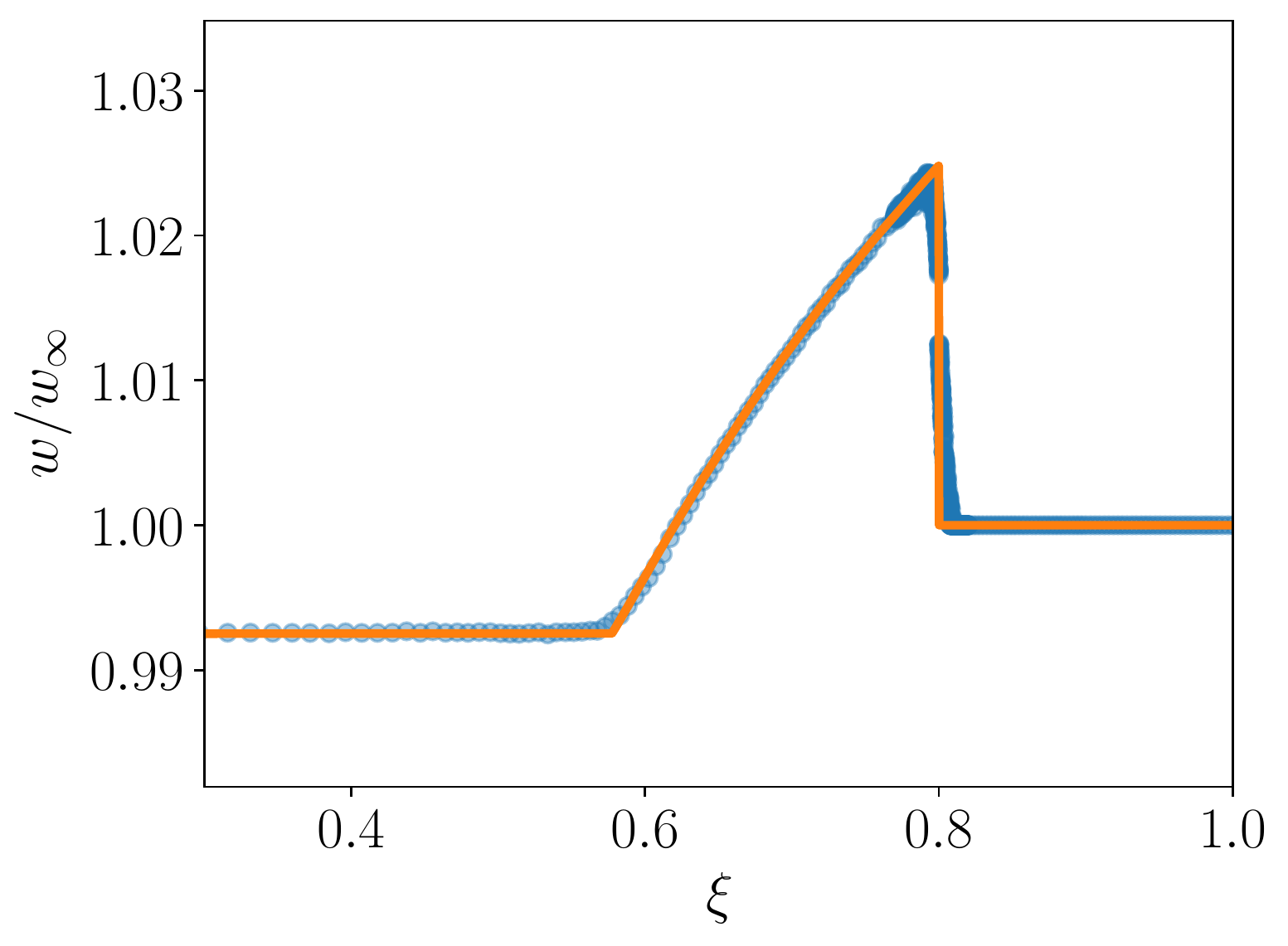} 
	\includegraphics[width=0.45\textwidth]{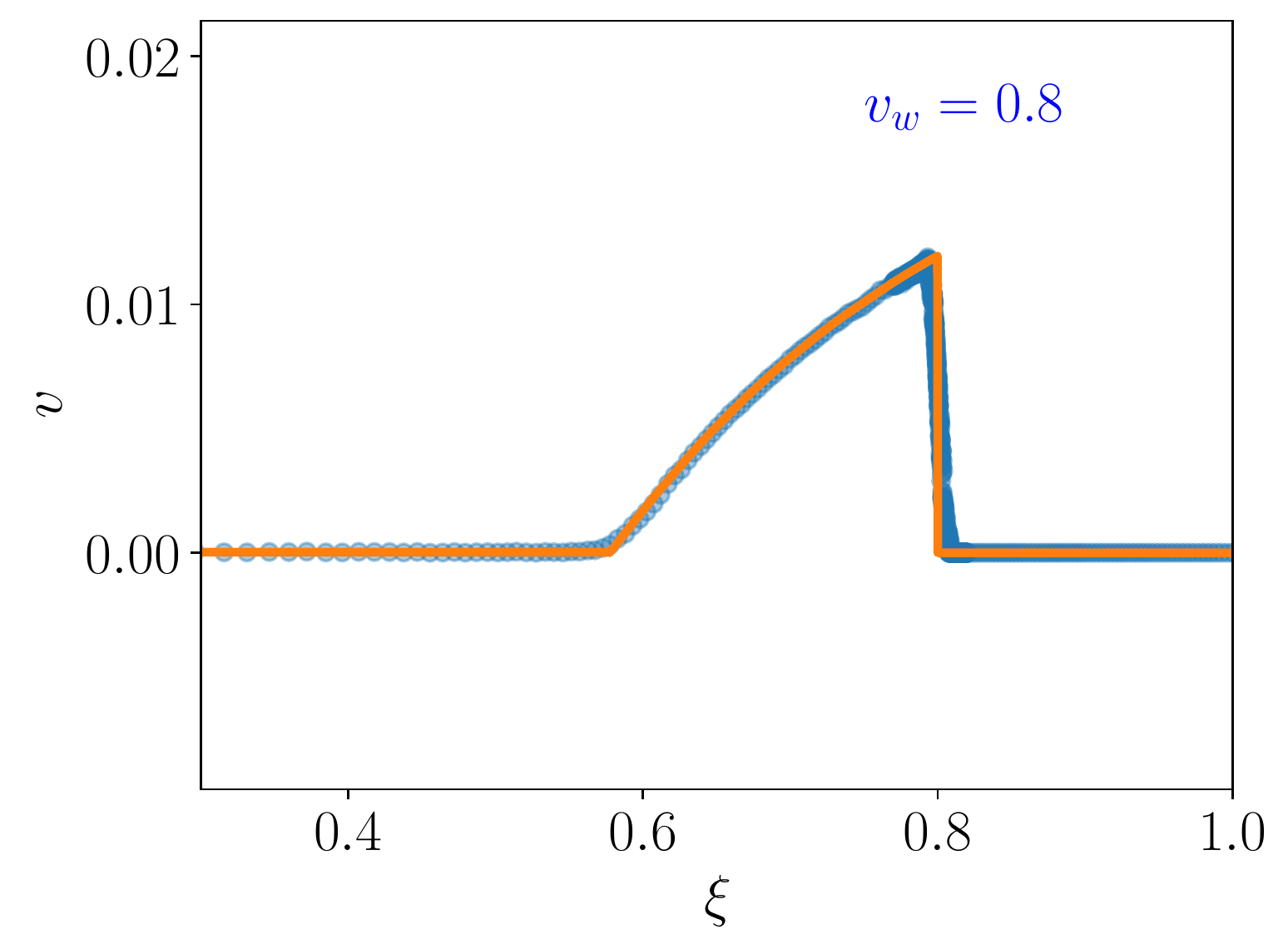} 
	\caption{
		Enthalpy $w$ (left) and fluid velocity $|\vec v|$ (right) averaged over a single bubble along the radius.
		The strength of the phase transition is weak, $\alpha = 0.0046$, and the wall velocities are $v_w = 0.4, 0.6, 0.8$ (deflagration, hybrid, detonation) from top to bottom.
		We simulate the bubble in a box of $L = 20 v_w / \beta$ and $N = 256$ from $t = 0$ to $t = 6.9 / \beta, 9.45 / \beta$ and $9.8 / \beta$, respectively for deflagrations, hybrids and detonation.
		The blue points are $100,000$ randomly chosen points, while the orange line is the asymptotic self-similar profile. 
	}
\label{fig:1d_Evolution_W}
\end{figure}

\begin{figure}
	\centering
	\includegraphics[width=0.45\textwidth]{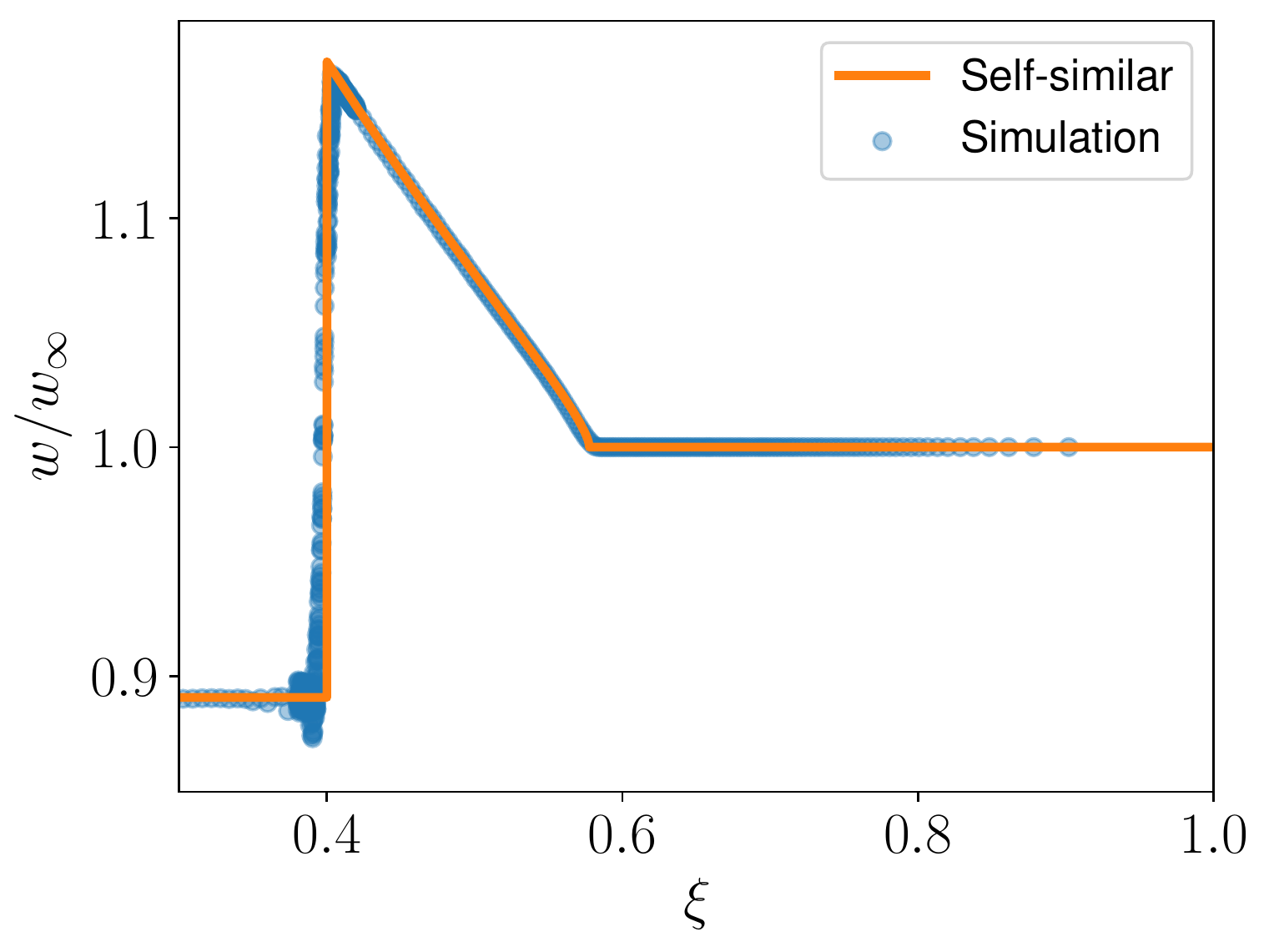}
	\includegraphics[width=0.45\textwidth]{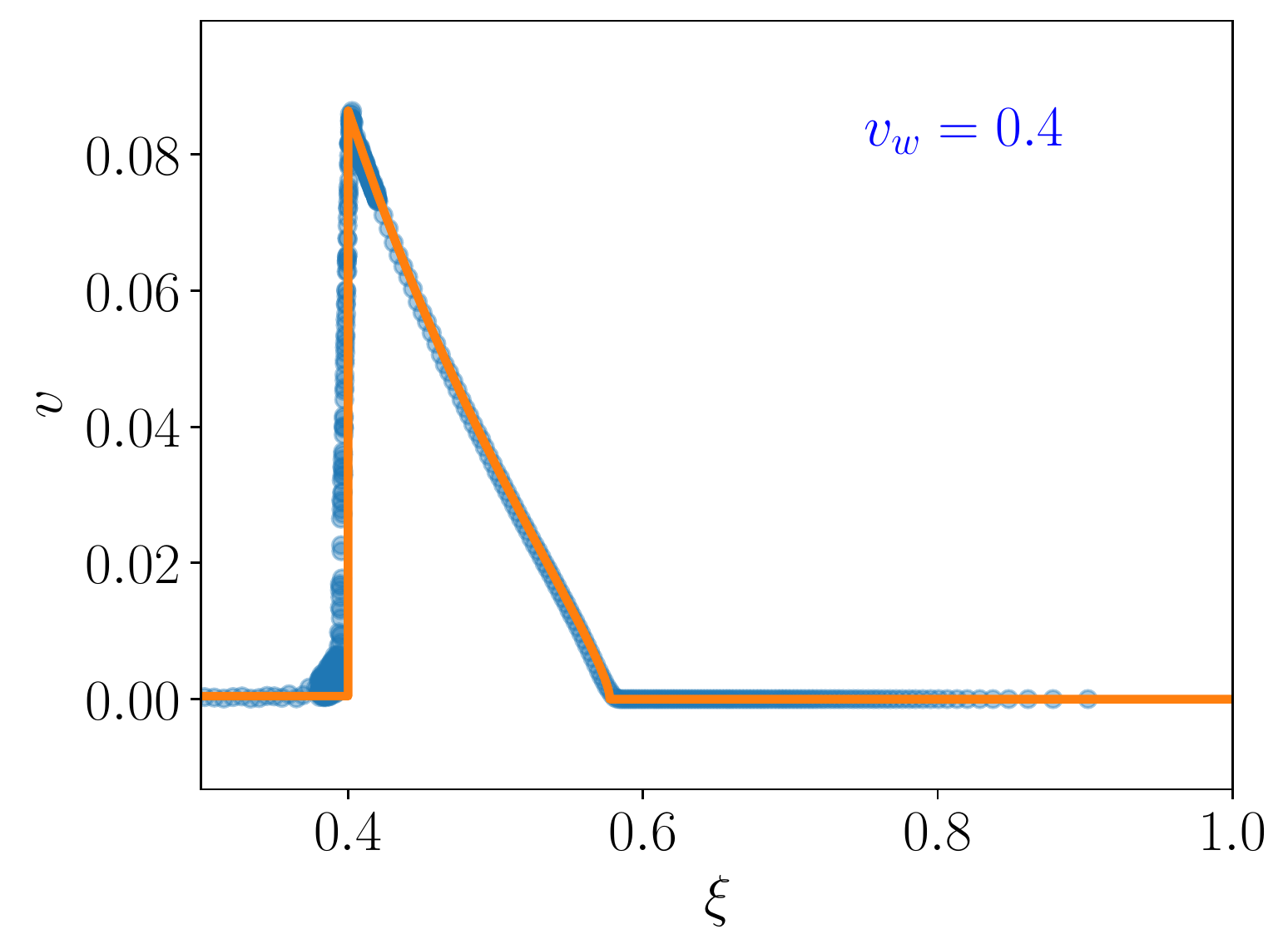}
	\includegraphics[width=0.45\textwidth]{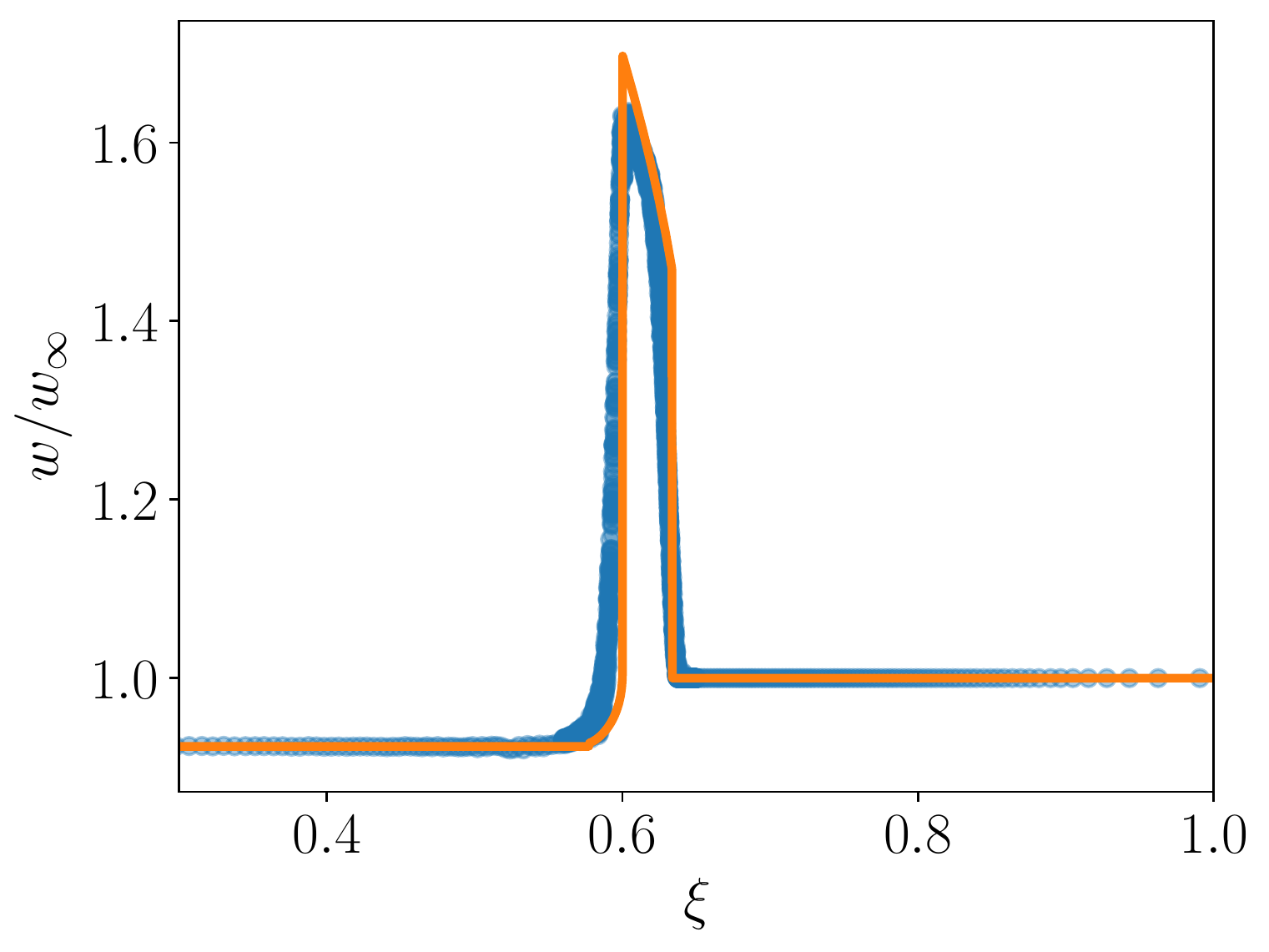}
	\includegraphics[width=0.45\textwidth]{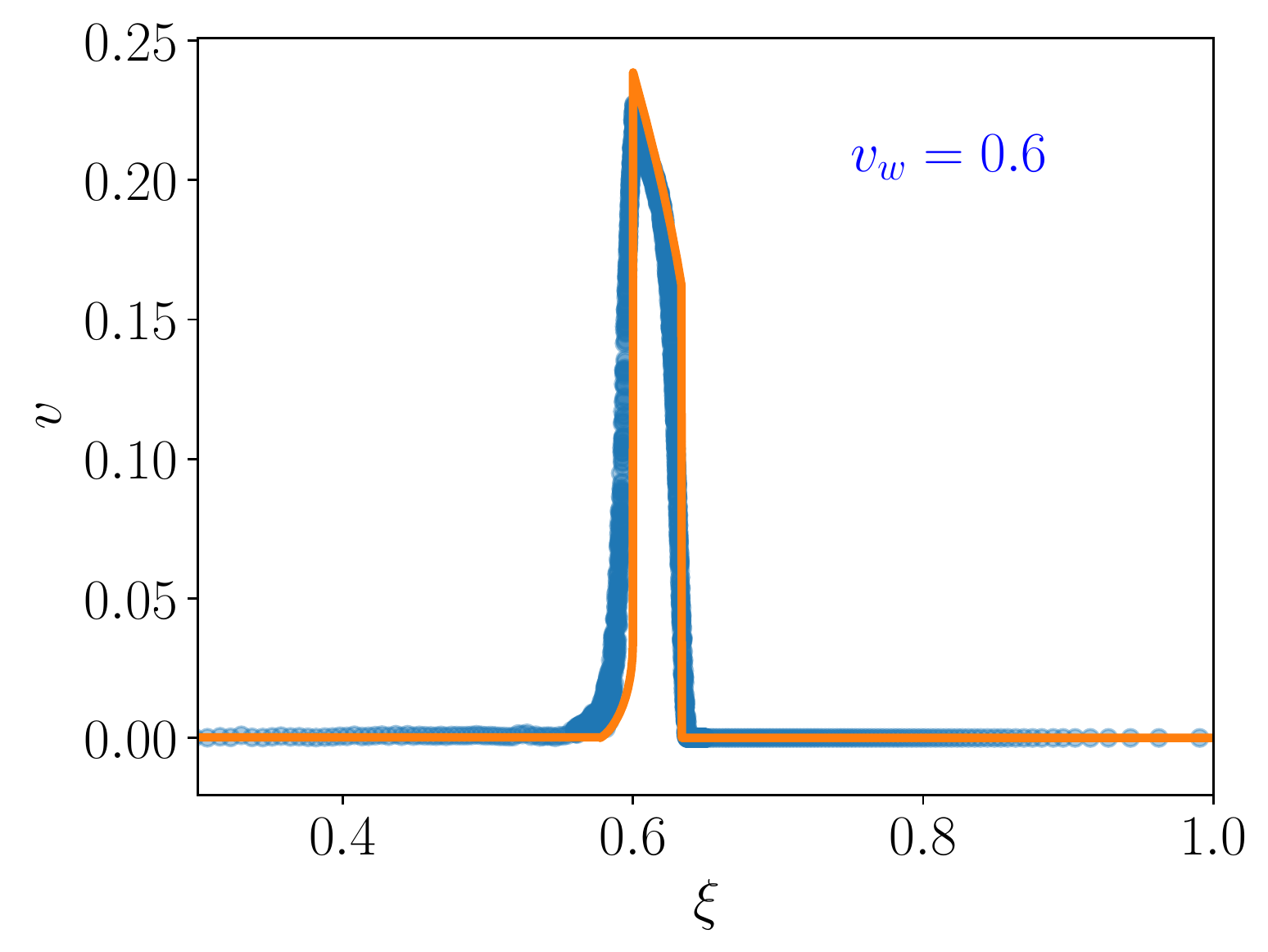}
	\includegraphics[width=0.45\textwidth]{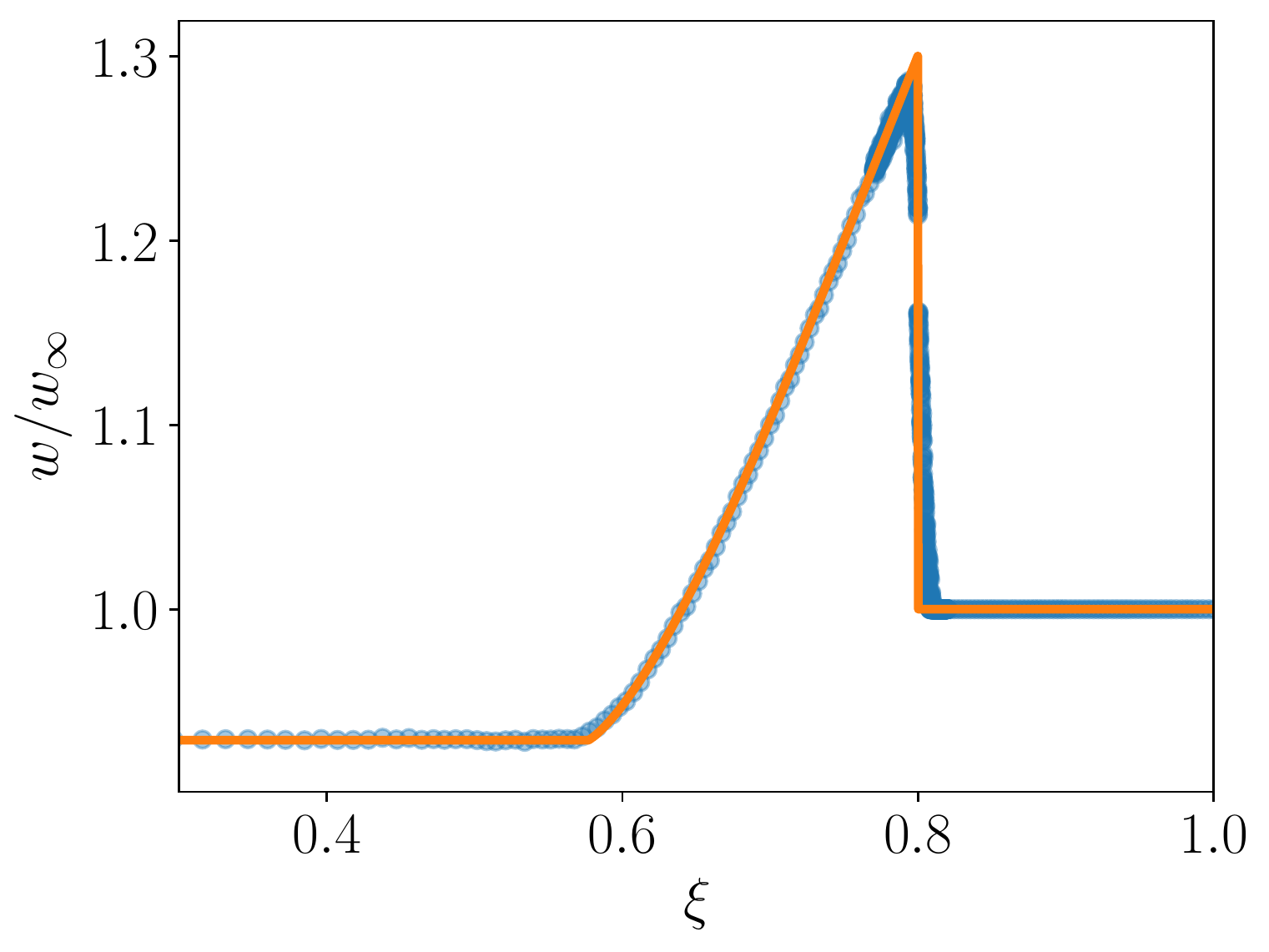}
	\includegraphics[width=0.45\textwidth]{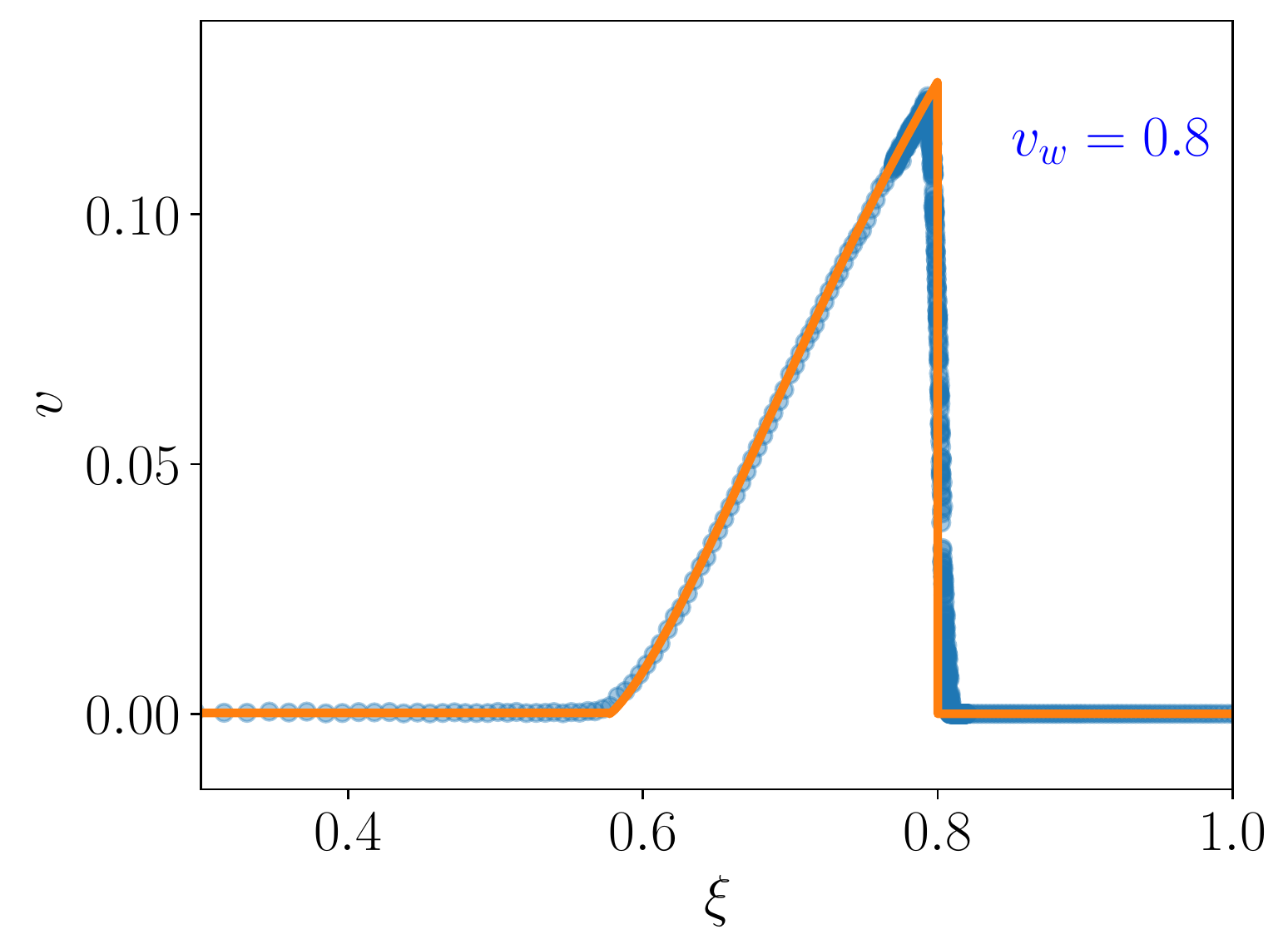}
	\caption{
		Same as Fig.~\ref{fig:1d_Evolution_W} for intermediate phase transitions, $\alpha = 0.05$. 	}
	\label{fig:1d_Evolution_I}
\end{figure}

We summarize the results for the single-bubble evolution of weak phase transitions ($\alpha = 0.0046$) in Fig.~\ref{fig:1d_Evolution_W} and intermediate PTs ($\alpha=0.05$) in Fig.~\ref{fig:1d_Evolution_I}. We show the self-similar solutions for the velocities and enthalpy for three types of profiles: deflagrations, hybrids (deflagrations with a rarefaction wave), and detonations. We use $L = 20 v_w / \beta$ and $N = 256$, evolving the profiles from $t = 0$ to $t = 6.9 / \beta, 9.45 / \beta$ and $9.8 / \beta$, respectively for deflagrations, hybrids and detonation\footnote{
Those times are chosen as the maximal times before the profiles hit the boundary of the box.
}.
For proper deflagrations ($v_w = 0.4$) and detonations ($v_w = 0.8$) the structures are reproduced quite well for weak transitions. For supersonic deflagrations with a rarefaction wave (so-called hybrid solutions),
the profiles are much thinner and the resolution of the grid is too small to properly resolve the shape of the fluid. In essence, we therefore expect that for thin shells the GW power is underestimated, and a clear separation of bubble size and sound shell thickness might be lacking. 
For intermediate transitions, deflagrations and detonations are well resolved with regards to large bubbles being self-similar at the time of collision, and the hybrid case performs better when compared to weak transitions since the shell thickness is larger.

Altogether, the 3D Higgsless simulation approach quickly converges to the correct single-bubble wall profile with outstanding precision and furthermore maintains an accurate time evolution of the shocks. The correct reproduction of the self-similar profile is somewhat independent of $\alpha$ and rather depends on the shell thickness. This indicates that pushing this framework to even stronger PTs would be straightforward with the cost of eventually adapting the time step of the simulation, as mentioned in Sec.~\ref{sec:KTscheme}.

We leave the study of stronger transitions to future work, and, for comparability with the literature \cite{Hindmarsh:2017gnf,Jinno:2020eqg}, restrict ourselves here to weak ($\alpha = 0.0046$) and intermediate ($\alpha = 0.05$) PTs.

\section{Numerical results}
\label{sec:res}

\begin{figure}[h]
	\centering
	\includegraphics[width=0.48\textwidth]{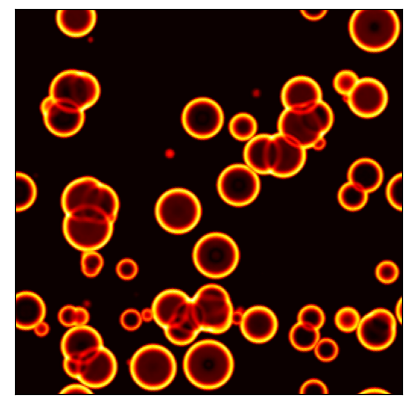} 	
	\includegraphics[width=0.48\textwidth]{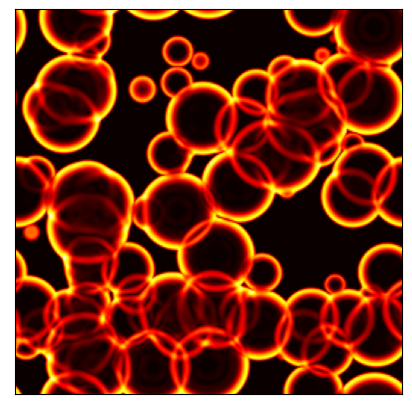} 	
	\includegraphics[width=0.48\textwidth]{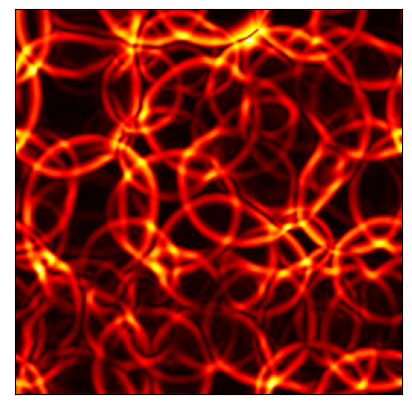} 
	\includegraphics[width=0.48\textwidth]{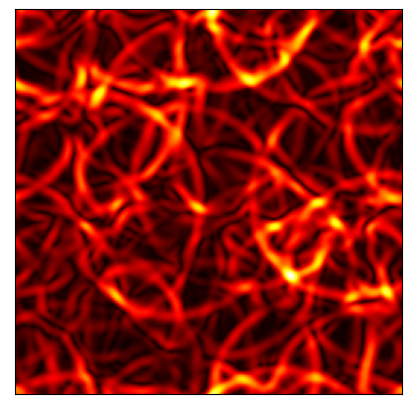} 
	\caption{ {Kinetic energy $v^2$ in different} simulation snapshots: $t = 2.7/\beta$ (top left),  $5.4/\beta$ (top right), $10.8/\beta$ (bottom left) and $ 20.1/\beta $ (bottom right). We use box size $L = 40v_w/\beta$, weak transitions and $v_w = 0.8$.}
	\label{fig:NLexampleHenrique}
\end{figure}

Before presenting our data on the GW spectrum produced during the phase transition, we present a few 2D time slices of the fluid kinetic energy distribution. Fig.~\ref{fig:NLexampleHenrique} shows the energy density at different time steps, for weak transitions and box size $L = 40v_w/\beta$. Initially, we note that the bubble size is the predominant physical scale. At later stages, we notice the emergence of a second scale, namely the sound shell thickness. Those are the two main scales that parametrize the GW spectra. Note however that at late times some imprints of the original bubbles are still visible.
The simulation slice we show is at the edge of the simulation volume while the first nucleated bubble is by default 
in the center of the simulation. Hence it is not visible in these slices and the bubble sizes seem quite homogeneous.

\begin{figure}
	\centering
	\includegraphics[width=0.6 \textwidth]{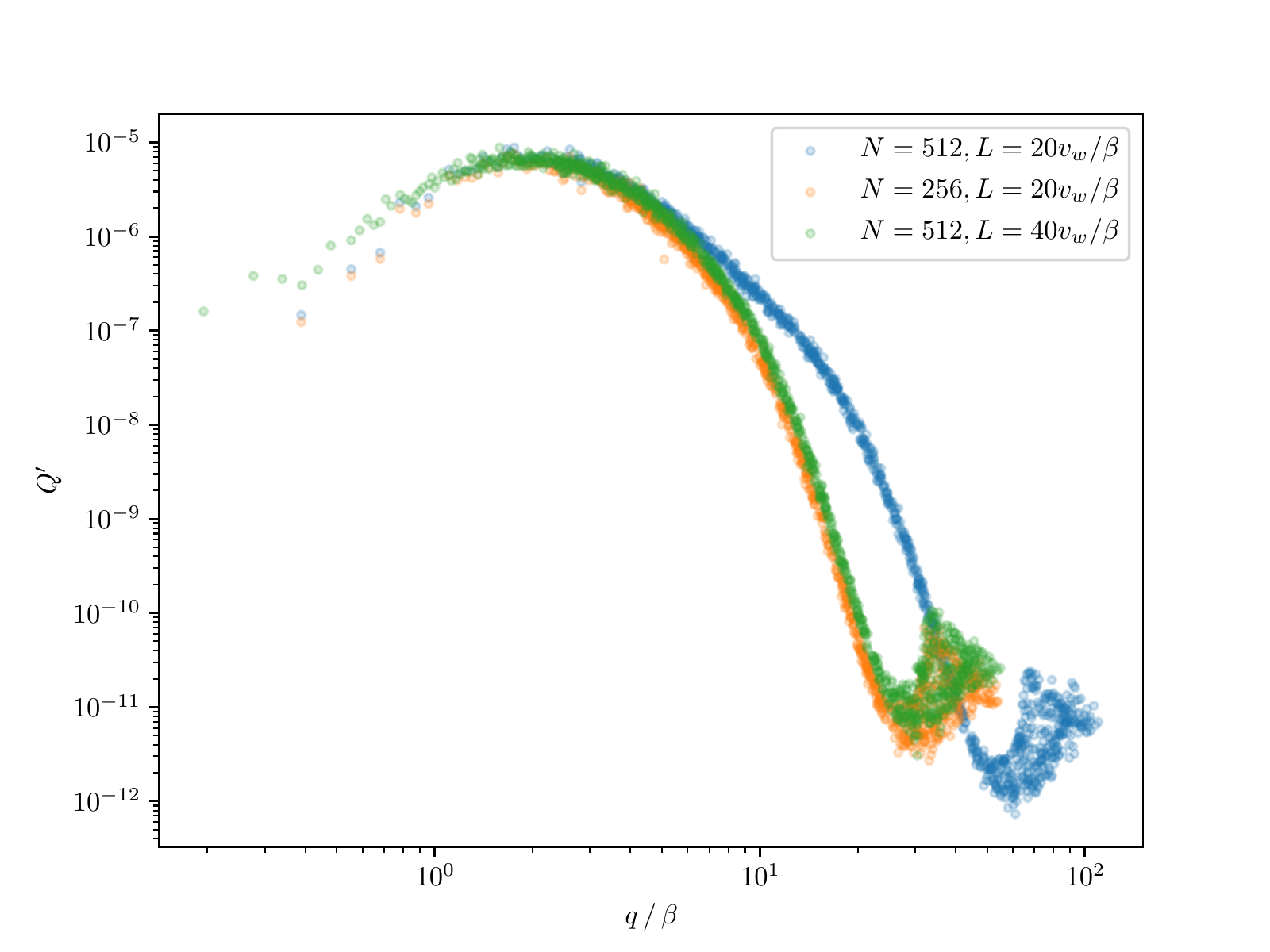} 
	\caption{Example of the impact of the box size $L$ and grid size $N$ on the resulting power spectrum.	We use $v_w = 0.8$ and intermediate strength ($\alpha = 0.05$).
	}
	\label{fig:NLexample}
\end{figure}

First, we will study the impact of the box size $L$ and the grid size $N$ on the 
resulting GW spectra. An example is given in Fig.~\ref{fig:NLexample}. We show box sizes $L=20v_w/\beta$ and $L=40v_w/\beta$ for 
$N=512$ and also one spectrum with $L=20v_w/\beta$ and $N=256$ (we vertically shifted the latter spectrum by a factor of $0.85$ to make it visible since otherwise the two sets of points overlap).
The number of bubbles is $N_b \sim 300$ and $\sim 2700$ for the small and large box, respectively.
The integration time in the Fourier transformation is for all simulations chosen to range from $16/\beta$ to $32/\beta$.
This integration range should make contributions from before percolation relatively small, while 
(at least for large box sizes) the first bubble does not have enough time to collide with its mirror images. 
Still, the softest modes are only tracked for a few oscillations in the Fourier transformation with 
respect to time, which can introduce systematic errors (see App.~\ref{sec:IRtail}). 
One can see that the spectrum from the smaller grid ($N=256$) agrees quite precisely with the result from 
the larger grid in the IR when the box size is the same.
Likewise, the spectrum from the smaller grid ($N=256$) agrees quite precisely with the result from the larger grid in the UV when grid spacing is the same. Depending on the parameters, the GW spectra
will display several features at different scales. The two relevant physical scales are the imprint of the bubble size (or PT duration) and the sound shell thickness. Moreover, the box size will constrain the available momenta in the IR while grid spacing and various sources of viscosity will lead to exponential damping in the UV.
A detailed discussion of this effect will be provided below.
Accordingly, different box sizes will facilitate the best measurements for the various physical observables. Also notice that the power spectrum is generally reduced by finite size effects in the IR and UV. The loss of power in the UV corresponds to a reduction in the average kinetic energy which we study in App.~\ref{sec:kinetic}.
Extrapolating to very large grid size, we estimate that this leads to a reduction of the momentum-integrated GW signal by about $20\%$.

\begin{figure}
	\centering
	\includegraphics[width=\textwidth]{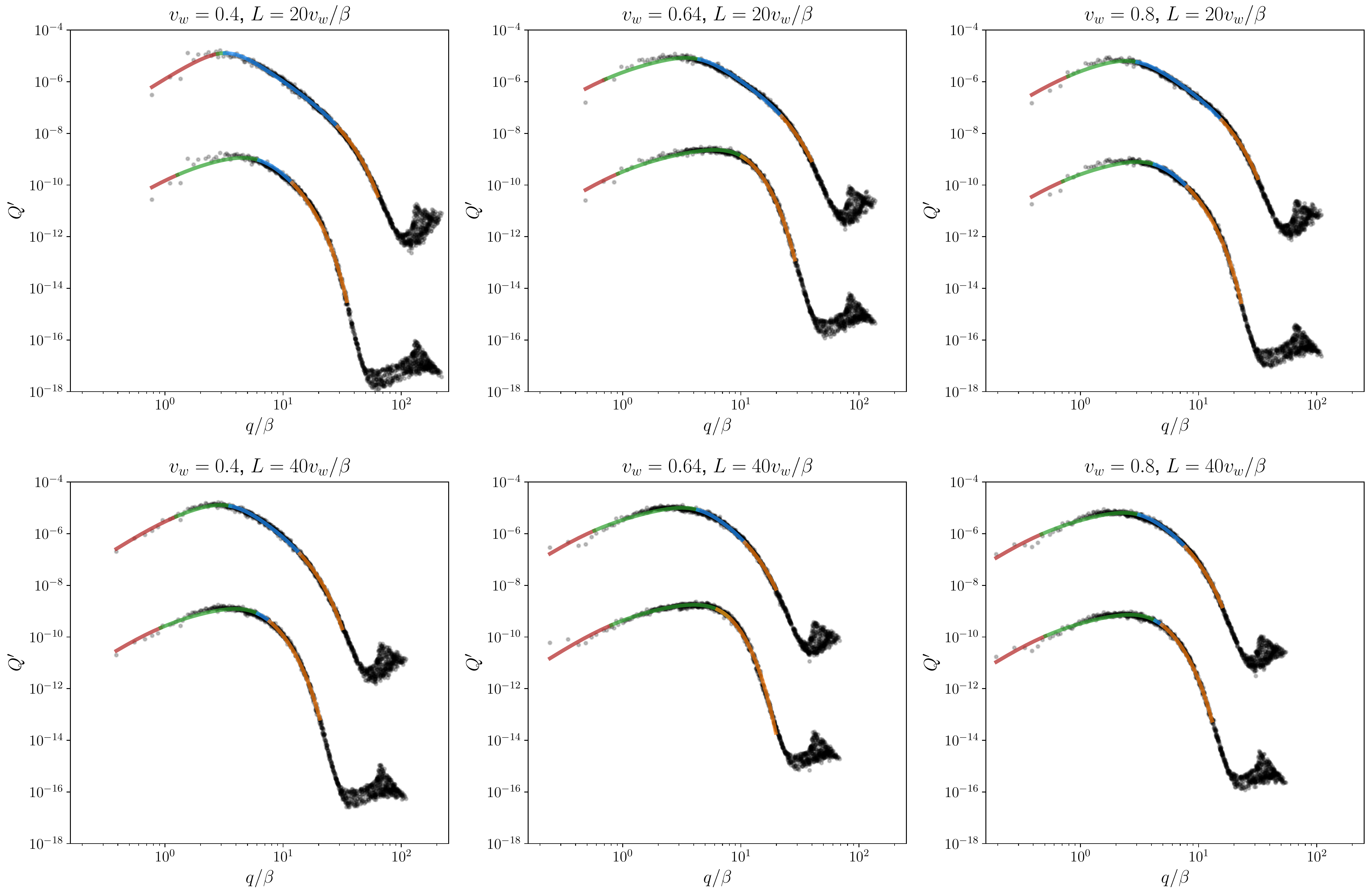} 
	\caption{
		Some example spectra for weak (lower lines) and intermediate (upper lines) PTs and $N=512$. The parameters are given in the captions. 
The colored lines are the shape function (\ref{eq:shape}), with different colors indicating the regions separated by $q_0$, $q_1$ and $q_e$.
	}
	\label{fig:spectraFit}
\end{figure}

Next, we study the slope of the power spectrum in the three regimes bounded by the bubble size and sound shell thickness. Fig.~\ref{fig:spectraFit} shows some example spectra. 
The parameters have been chosen in order to 
showcase the different asymptotic behaviors. For large boxes, the infrared 
behavior can be most easily determined (bottom panels). For weak phase transitions and 
wall velocities close to the Jouguet velocity, the plateau between bubble size and 
shell thickness is most visible (middle panels). For small box size and generic wall velocities,
the UV tail is most visible. Exponential damping from numerical viscosity 
seems to be stronger for weaker phase transitions (lower lines).

In principle, the spectrum that we observe is a double-broken power law. At small momenta, the 
spectrum increases as $q^3$ up to the scale that corresponds to the bubble size
(see App.~\ref{sec:IRtail} for a more detailed discussion). 
Between the momentum corresponding to the bubble size and the momentum corresponding to the thickness of the 
sound shells, the spectrum seems to increase as $q$. Next, for large momenta, the spectrum tends to 
decrease as $q^{-3}$. We indicated these power laws in the different spectra whenever they are visible. 
Finally, for very large momenta the power spectrum is suppressed exponentially. 

Given the clear scaling of the GW spectra in the IR, intermediate and UV regimes, we fix the exponents in the shape function, contrary to what was done in the hybrid simulations~\cite{Jinno:2020eqg}. It allows for a more precise extraction of the peak and knee positions.
We fit the spectra using the following shape function
\be
Q'(q) = Q_{\rm int}' \times S_f (q) \, ,
\ee
where
\be
S_f(q) = S_0 \times \frac{(q/q_0)^3}{1 + (q/q_0)^2 [ 1 + (q/q_1)^4 ]} \times e^{-(q/q_e)^2} \, .
\label{eq:shape}
\ee
We normalize the shape function using $S_0$ such that $\int d \ln q \, S_f(q) = 1$ and $\int d \ln q \, Q'(q) = Q_{\rm int}'$.
The exponential suppression is chosen such that it fits well with the observed spectra.

We show in Fig.~\ref{fig:spectraFit} the fits of the shape function to our example spectra (for $N=512$).
Overall, the shape function accurately captures the behavior of the measured spectra.
For the smaller box size, the first knee $q_0$ might be measured somewhat poorly and 
quite generally the fit provides a too large value for $q_0$.  
Apparently, there are just not enough momenta in the IR part of the spectrum. 
Likewise, for 
an extended plateau, $q_1 \gg q_0$, the exponential decay in the UV might inhibit a 
determination of the scale $q_1$, as seen for the weak phase transition in the top middle panel. 

\begin{figure}
	\centering
	\includegraphics[width=\textwidth]{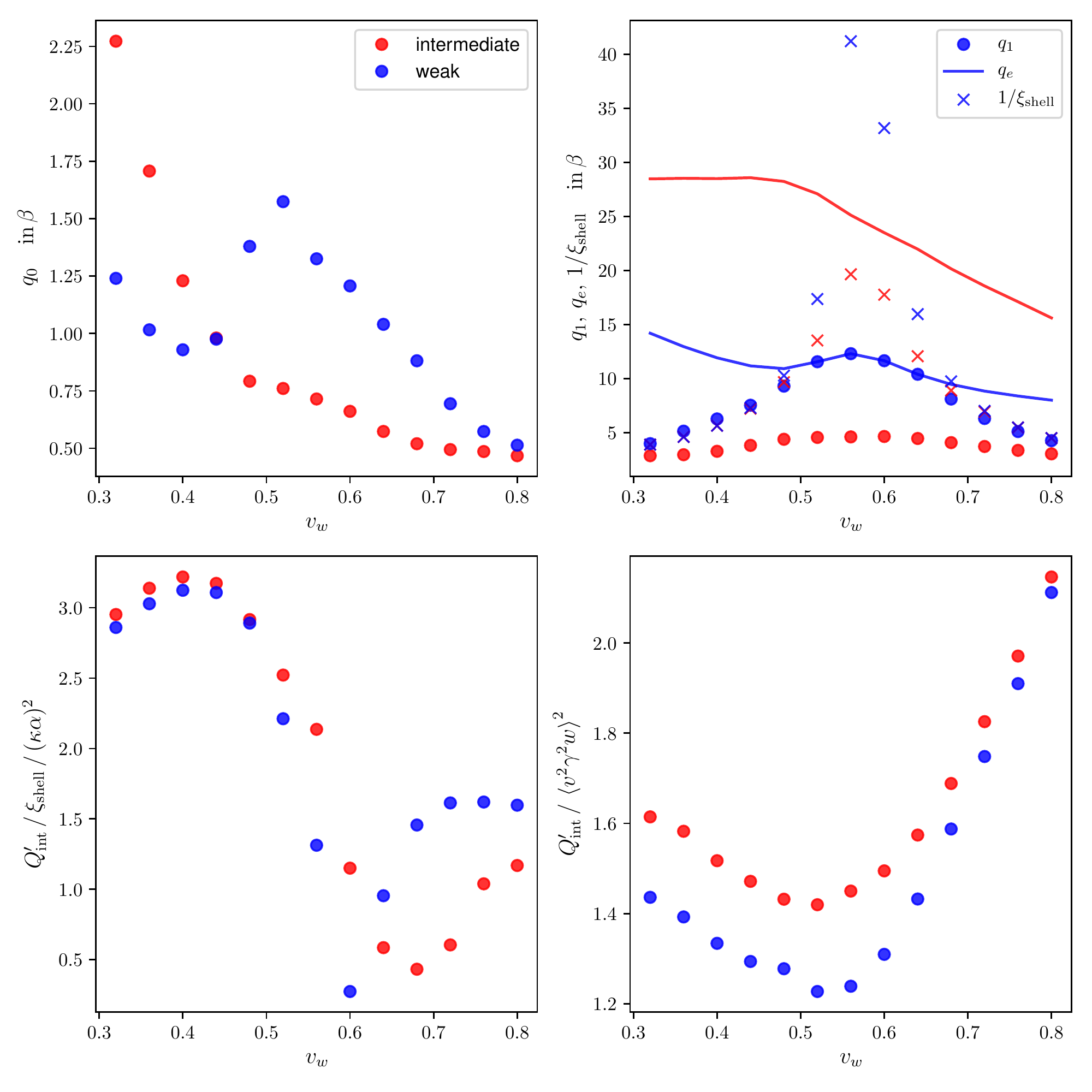} 
	\caption{
		The extracted fitting parameters as functions of the wall velocity. Blue (red) points correspond to phase transitions with weak (intermediate) strength and $\alpha = 0.0046$ ($\alpha = 0.05$). The top panels show the scaling of the peak and knee positions as a function of $v_w$. In the left panel we show the IR knee position $q_0$. In the right panel we show the UV peak $q_1$ (dots), the shell thickness (crosses) defined in Eq.~(\ref{eq:xi_shell}) and the exponential damping $q_e$ (solid lines). The bottom panels show the integral of the GW spectra $Q'_{\rm int}$ over momenta defined in Eq.~(\ref{eq:Q_int}) normalized by the $\xi_{\rm shell}(\kappa\alpha)^2$ (left) and by the kinetic energy squared $\langle v^2 \gamma^2 w \rangle^2$ measured in the lattice (right).}
	\label{fig:params}
\end{figure}

Motivated by these findings, we will in the following use simulations with larger boxes ($L = 40v_w/\beta$) to measure the IR quantity $q_0$ and simulations with smaller boxes ($L = 20v_w/\beta$) to measure $q_1$ and $q_e$. 
Figure~\ref{fig:params} shows our final results for $N=512$. The full data is given in App.~\ref{sec:full}.

\vskip 0.5 cm

The top left panel of Fig.~\ref{fig:params} shows the scale $q_0$ related to the bubble size (or to the duration of the phase transition).
As mentioned before these measurements are obtained in a simulation with a relatively large box size, $L=40v_w/\beta$, and 
relatively many bubbles, $N_b \gtrsim 2000$. Using a large enough box size is essential for this measurement and simulations with fewer bubbles and smaller boxes tend to overestimate $q_0$ (see App.~\ref{sec:full}). Notice that for the phase transitions with 
intermediate strength, one can observe a clear downward trend, while for the phase transitions with weak strength, there 
is a feature close to the speed of sound. 
Our data indicates that the IR knee has more complex behavior than previously observed.

The top right panel of Fig.~\ref{fig:params} shows several UV quantities
and all measurements have been obtained in simulations with small box size $L=20v_w/\beta$: The scale $q_1$ -- that is related to the sound shell thickness --
the onset of exponential damping $q_e$ and the inverse shell thickness $1/\xi_{\rm shell}$ as measured in self-similar spherically symmetric solutions to the hydrodynamic equations.
The shell thickness is defined as
\begin{align} \label{eq:xi_shell}
\xi_{\rm shell} &:= \xi_{\rm front} - \xi_{\rm rear},
\end{align}
where\footnote{
Concerning the definition of $\xi_{\rm shell}$, there is a typo for the hybrid case in the text of \cite{Jinno:2020eqg}, while in the analysis the correct expression was used.
}
\begin{align}
\xi_{\rm front} &= \left\{
\begin{matrix}
\xi_{\rm shock} &\quad {\rm (deflagration,~hybrid)}
\\[0.2cm]
\xi_w &\quad {\rm (detonation)}
\end{matrix}
\right. ,
\qquad
\xi_{\rm rear} = \left\{
\begin{matrix}
\xi_w &\quad {\rm (deflagration)}
\\[0.2cm]
c_s &\quad {\rm (detonation,~hybrid)}
\end{matrix}
\right. .
\end{align}
The shell thickness $\xi_{\rm shell}$ is determined as described in~\cite{Jinno:2020eqg}, while the frequencies $q_1$ and $q_e$ are extracted from the GW spectrum.
In our fit, we force $q_1$ to lie between 
$q_0$ and $q_e$. Therefore, when $q_0$ approaches $q_e$, the exponential damping inhibits an accurate measurement of the scale $q_1$. This happens in our case when the wall velocity is close to the speed of sound and the phase transition is not too strong.
For weak phase transitions, the proposal $1/\xi_{\rm shell}$ from \cite{Jinno:2020eqg} seems to fit $q_1$ quite well, while $q_1$ is significantly smaller for stronger phase transitions. This might be due to nonlinear effects that are more important for 
stronger phase transitions and tend to wash out any sharp features. Another curious observation is that $q_e$ is significantly lower for weak phase transitions than for stronger ones. This means that the scale $q_e$ cannot be directly related to the 
grid spacing in both cases, since the grid spacing is the same in both. 
The difference can possibly be attributed to numerical viscosity; however, a contribution from physical dynamics remains an exciting possibility since faster-moving plasma configurations form shocks faster.
In the spherical simulations we performed (see Figs.~\ref{fig:1d_Evolution_W} and \ref{fig:1d_Evolution_I}) we could not observe
any big difference in the damping scales which also supports that the exponential cutoff might be physical and result from the dynamics after percolation completed.
One might think that the reason for the rise in $q_e$ around $v_w \sim c_s$ is the imposed condition that $q_1 < q_e$. However, relaxing this condition yields very similar results for all but one point where $q_1$ is only marginally larger than $q_e$. The condition $q_1<q_e$ can therefore not alone be responsible for the rise in $q_e$. 

The bottom left panel of Fig.~\ref{fig:params} shows the amplitude of the GW spectrum. To be precise, we show the integral 
\be \label{eq:Q_int}
Q^\prime_{\rm int} := \int d \ln q \, Q^\prime \, , 
\ee
divided by the shell thickness $\xi_{\rm shell}$ and the efficiency to convert the latent heat into kinetic energy $(\kappa \alpha)^2$, as advocated in~\cite{Jinno:2020eqg}. Our results overall resemble very well the findings in~\cite{Jinno:2020eqg}.
The amplitude changes by about a factor of 10 across different phase transitions and wall velocities. Part of this large spread 
is that the conversion factor $\kappa \alpha$, that is measured in self-similar spherically symmetric solutions, overestimates the
conversion of latent heat into kinetic motion throughout the phase transition when the wall velocity is close to the speed of sound~\cite{Jinno:2020eqg}. 
A large part of this effect is compensated by the additional factor $\xi_{\rm shell}$ that we introduced in the normalization but a rather large 
spread in values still remains. Recall that for weak phase transitions, the simulation 
cannot properly resolve the fluid profile for wall velocities close to the speed of sound, and these parameter points should be taken with a grain of salt.

A better normalization is 
given in terms of the kinetic energy measured in the full 3D simulation, as shown in the 
bottom right panel of Fig.~\ref{fig:params}. In this case, no additional factor $\xi_{\rm shell}$ has to be fudged in and the normalization is surprisingly independent of the wall velocity 
or the strength of the phase transition (within a factor $\sim 2$). Unfortunately, the kinetic energy in the 
fluid is hard to come by without running the full simulation. All in all, building a simulation template bank for 
GW spectra from PTs that captures nonlinearities might be necessary to constrain LISA parameter space, 
similar to what is done in the context of LIGO for GR simulations.

\section{Discussion and conclusions}
\label{sec:disc}

In this section we condense different results available in the literature, discussing differences and agreements between these approaches in the different regimes of the GW spectra. 

First, we shortly describe the models that focus on the scalar contribution. Our simulations are dominated by the
contributions from the propagating sound shells and do not contain any scalar field. This part, therefore, is presented rather for completeness than for a comparison with our results. The approaches that aim at quantifying the contribution from the 
scalar field, or more generally from the collision of bubbles, are

\begin{itemize}
	\item {\bf Envelope}: Considers only the scalar field. It approximates the stress-energy tensor of the scalar field by a thin layer around the bubble regions that have not yet collided. It is important to highlight that what is often referred to as envelope approximation embraces two different approximations \cite{Kosowsky:1992rz}: first that the energy density is confined in a {\it thin wall}, second that only the uncollided regions contribute to the stress-energy tensor. The envelope contribution was analytically solved by \cite{Jinno:2016vai} after the numerical simulation results of \cite{Huber:2008hg}. 

	\item {\bf Bulk flow}: Extends the envelope approximation by lifting the approximation that the collided regions do not contribute to the stress-energy tensor. Instead of instantaneous dissipating, it considers that the collided part slowly decays. Analytically solved by \cite{Jinno:2017fby} and subsequently had its scaling confirmed by numerical simulations \cite{Konstandin:2017sat}.
	
	\item  {\bf Scalar field in the lattice}: Lattice simulation of the scalar field alone, no interaction with the underlying fluid \cite{Cutting:2018tjt}. 
\end{itemize}

We sum up in Table~\ref{table:slopes_scalar} the different scalings observed by those different models for the scalar part. Note that there is only one scale, related to the bubble size (or PT duration), dividing the IR and the UV parts (the Higgs scale is very deep in the UV). Note that both the envelope and the scalar lattice agree on the IR $q^3$ scaling, supported by causality \cite{Caprini:2009fx}. Bulk flow modeling of the IR captures a longer-lasting source of GWs, leading to a $q^1$ scaling. In the UV, bulk flow differs by having fewer kinks and cusps in the bubble configurations compared to the envelope approximation, which leads to a faster decay of the spectrum in the UV, in particular for non-relativistic walls. Lattice simulations also observe a stronger decay. The position of the peak is slightly different between the envelope approximation and the lattice \cite{Cutting:2018tjt}. 

\begin{table}[t!h]
	\centering
	\begin{tabular}{|c||c|c|c|} \hline 
		&  IR &  UV &  References\\ \hline\hline  
		
		Envelope & $3$ & $-1$  & \cite{Huber:2008hg,Jinno:2016vai}  \\ \hline
		
		Bulk flow  &  $1$ & $-3$ &  \cite{Jinno:2017fby,Konstandin:2017sat}   \\ \hline 
		
		Scalar lattice  &  $3$& $-1.5$ & \cite{Cutting:2018tjt} \\ \hline 
	\end{tabular}
	\caption{\small Slopes of the GW spectrum for the models that describe the scalar field. }
	\label{table:slopes_scalar}
\end{table}

Since the observation that bubbles hardly {\it run away} \cite{Bodeker:2017cim} and that sound waves in the plasma last for longer times \cite{Hindmarsh:2013xza}, the fluid dynamics has been demonstrated to be the dominant contribution to the GW spectrum\footnote{
Note that, if the walls are highly relativistic, the fluid dynamics might be somewhat more subtle.
}. The models that describe the plasma evolution and the resulting gravitational waves are

\begin{itemize}
	\item {\bf Sound-shell approximation}: The sound-shell approximation is an analytical method that assumes that sound waves freely propagate and overlap \cite{Hindmarsh:2016lnk,Hindmarsh:2019phv}. This overlap assumes linearity of the velocities. The final GW spectrum calculation also assumes that the velocity field is Gaussian at late times and has a certain correlation in time, such that the four-point function for the velocity (that enters the GW spectrum) can be written in terms of the velocity power spectrum.
	\item {\bf Scalar field + fluid lattice simulations}: Simulation of the scalar field coupled with the fluid via a phenomenological friction term \cite{Hindmarsh:2013xza,Hindmarsh:2015qta,Hindmarsh:2017gnf,Cutting:2019zws}. For the scalar field, the parametric dependence of the phenomenological friction term on temperature and scalar field as well as a specific free energy is assumed. 
	\item {\bf Hybrid simulations}: The hybrid scheme of \cite{Jinno:2020eqg} models spherically symmetric bubbles before and 
after collision and then embeds the corresponding enthalpy and fluid velocity profiles into a 3D grid according to an exponential bubble nucleation history. The spherical simulations include the Higgs field only as a boundary condition, analogously to the 3D implementation in this work. Still, the embedding assumes linearity since the fluid is 
constructed as a superposition of collided and uncollided bubble shell fragments.
	\item {\bf Higgsless simulations}: The approach in this work. The Higgs field only enters as a boundary condition in the equation of 
state (see Sec.~\ref{sec:fluid}) and the hydrodynamic equations are solved fully nonlinearly.
\end{itemize}

\begin{table}[t!h]
	\centering
	\begin{tabular}{|c||c|c|c|c|} \hline 
		&  IR &  Intermediate &  UV &  References\\ \hline\hline  
		
		Sound shell  &  $9$ &  $1$ & $-3$ & \cite{Hindmarsh:2016lnk,Hindmarsh:2019phv}  \\ \hline 
		
		Scalar + fluid lattice&  - & $1$& $-3$ & \cite{Hindmarsh:2013xza,Hindmarsh:2015qta,Hindmarsh:2017gnf,Cutting:2019zws}  \\ \hline
		
		Hybrid  &  [$2$,$4$] & [$-1$,$0$] & [$-4$,$-3$] &  \cite{Jinno:2020eqg} \\ \hline

		Higgsless  &  $3$ & $1$ & $-3$ &  This work \\ \hline 
		
	\end{tabular}
	\caption{\small Slopes of the GW spectrum for the plasma sound waves. Those models include features at $q_0$ and $q_1$ separated by three slopes (IR, intermediate, and UV). }
	\label{table:slopes_fluid}
\end{table}

We display in Table~\ref{table:slopes_fluid} the results for the slopes that constitute the GW power spectrum. One immediate observation is that the sound shell approximation of \cite{Hindmarsh:2016lnk,Hindmarsh:2019phv} predicts an IR behavior that is quite different than the one seen in simulations. The hybrid and Higgsless 
approaches are consistent with a $q^3$ behavior and also the scalar+fluid simulations disfavor a very 
steep power spectrum as seen by the sound shell approximation. Thus, even though a more analytic approach to predicting 
GW spectra is preferable, we will not compare our results with this method in detail. 

Likewise, the hybrid method of \cite{Jinno:2020eqg} has many limitations but no clear advantages compared to the scheme in the present work. 
The hybrid method is also Higgsless in the same sense as the scheme presented here, but it comes with more assumptions and limitations. It consists of linear superpositions of 1D profiles, while the present scheme is fully nonlinear. It only captures part of the fluid interactions by embedding the results from spherical simulations into a 3D grid, which will lead to artifacts that will surely increase over simulation time.
Still, there are some lessons to learn from comparing the Higgsless with the hybrid approach. One curious result that 
both approaches share is that for weak phase transitions the scale $q_1$ seems to follow the shell thickness $1/\xi_{\rm shell}$ closely, while for stronger phase transitions it does not. Hence, this difference cannot come from late-time nonlinear effects (since they are not accounted for in the hybrid approach). The most plausible explanation is that it comes from nonlinear effects 
during the first collision of sound shells (that are treated spherically symmetric but nonlinearly in the hybrid approach).
Likewise, the features seen in the amplitude of the spectrum are quite similar for hybrid and Higgsless simulations, 
which is reassuring (the amplitudes from the hybrid method are however larger by a factor of $2-3$, see the comments below).

Finally, let us compare the present approach with the scalar+fluid simulations of \cite{Hindmarsh:2013xza,Hindmarsh:2015qta,Hindmarsh:2017gnf,Cutting:2019zws}, which are the most accurate and 
sophisticated predictions of the GW spectrum so far. First, we would like to note 
that there are no large disparities between the two methods for most of the parameter space. Although the IR was not directly extracted from the GW spectrum in \cite{Hindmarsh:2017gnf} the intermediate and UV slopes 
 that constitute the GW power spectrum are consistent and also the amplitude is in rather good agreement. 
We compared individual results and the amplitude between 
the two methods typically differed by less than a factor of 2~\footnote{We compared the peak of the spectrum with the logarithmic plots in~\cite{Hindmarsh:2017gnf}.}. At the same time, we see 
many features in the data that have not been studied systematically before. The 
frequency of the knee, $q_0$, has a complex wall velocity dependence. The peak frequency $q_1$
 seems to follow $1/\xi_{\rm shell}$ for weak phase transitions and is systematically smaller
 when the phase transition is stronger.

The most important differences we identified between the Higgsless and the scalar+fluid simulations are 
the reduction in GW power for deflagrations as reported in~\cite{Cutting:2019zws} and later explored in \cite{Cutting:2022zgd}. The reduction is 
based on the fact that if the plasma is heated in front of the wall, the pressure difference that 
drives the expansion of the wall drops, and the wall velocity is reduced. In extreme cases, the bubble wall 
might completely stop which leads to an era of phase co-existence. 
But observing a slowdown of the bubble wall is clearly not possible in our Higgsless
scheme where bubbles are assumed to expand with a constant wall velocity. 
To a certain extent, this is a deficiency of the Higgsless approach that ignores the dynamics of the Higgs. However,
we would like to emphasize that any effect of this sort necessarily introduces a dependence on the modeling of friction into the 
framework. For example, the friction term used in the scalar+fluid simulations is phenomenological and represented 
by one additional term in the Higgs equation. This term has many unphysical properties: it is not correct in the relativistic limit, where the friction is known to plateau~\cite{Bodeker:2009qy}; it does not capture the dependence on the wall velocity close to the 
speed of sound (when the expansion mode transitions from a deflagration to a detonation); it does not capture the dependence of the friction on the Higgs VEV in the broken phase (that e.g.~leads to the curious effect that stronger phase transitions actually do not 
lead to faster walls in the SM~\cite{Moore:1995si}); and it does not have the proper temperature dependence. 
Likewise, the free energy used in the scalar+fluid simulations introduces a model dependence. For example, in the most recent simulations of~\cite{Cutting:2019zws}, a free energy was used that contains terms of order $T^4 \phi^2$ and $T^4 \phi^3$.
Even though this choice mimics the bag equation of state in both phases and seems to make the simulations more numerically robust,
this kind of free energy is hardly encountered in a concrete particle physics model. To what extent the reduction in GW power hinges on these assumptions is in our view an open question.

\vskip 1cm

In summary, Higgsless simulations are a novel and very efficient way to obtain the GW spectra from cosmological 
first-order phase transitions. The main assumptions are the bag equation of state (which can be relaxed at the 
cost of a more involved determination of $T^{ij}(K^\mu)$) and a constant expansion velocity of the bubble walls. 
The approach does not capture the dynamics of the Higgs field. The Higgs dynamics might be relevant
but also introduces a model-dependence, which makes the results less universal. One of our main results is that a robust measurement of the IR tail of the GW spectrum requires quite large box size and thousands of bubbles in the simulation. We observe the scaling $q^3$ for the IR tail of the spectrum and a quite rich behavior for the position of the first knee $q_0$ depending on the wall velocity and the PT strength. A recipe how to obtain today's observed GW spectra is given in App.~\ref{sec:spectra}.

The Kurganov-Tadmor scheme seems to be well suited to discretize the relativistic hydrodynamic PDE but other methods based on Riemann solvers should work equally well. Our simulations have been performed with very modest resources\footnote{Depending on the parameters, a single $N = 512$ simulation takes about 24 hours on a machine with 20 cores.} showcasing the efficiency of the Higgsless approach. 

In the future, we aim to explore different regimes of PTs. The 3D Higgsless approach is fully nonlinear and our numerical scheme should be able to explore strong phase transitions. Also, the Higgsless approach might be useful to study turbulence that is generated in the bubble collisions or the subsequent fluid dynamics, as studied previously in~\cite{Auclair:2022jod}.

\section*{Acknowledgements}
We would like to thank Mark Hindmarsh and Jorinde van de Vis for valuable comments on the draft. 
TK and IS are supported by the Deutsche Forschungsgemeinschaft (DFG, German Research Foundation) under Germany’s Excellence Strategy -- EXC 2121 “Quantum Universe” -- 390833306. 
HR is supported by the Excellence Cluster ORIGINS, that is funded by the Deutsche Forschungsgemeinschaft (DFG, German Research Foundation) under Germany’s Excellence Strategy - EXC-2094 - 390783311.
The work of RJ is supported by the grants IFT Centro de Excelencia Severo Ochoa SEV-2016-0597, CEX2020-001007-S, and by PID2019-110058GB-C22 funded by MCIN/AEI/10.13039/501100011033 and by ERDF.

\appendix

\section{Obtaining redshifted GW spectra}
\label{sec:spectra}

The main focus of our work is the technical implementation of the GW simulations and the 
spectra at production time. The final observed spectra are obtained by redshifting these 
results. In this appendix, we will briefly review how to obtain today's observed spectra.

First, for a specific model, the strength parameter $\alpha$, the duration of the phase transtions
$\beta/H$, the wall velocity $v_w$, and the phase transition temperature $T$ have to be obtained (see~\cite{Caprini:2019egz} and \cite{Giese:2020rtr, Giese:2020znk} for recent summaries on these topics). Next, from these 
quantities the properties of spherically-symmetric profiles such as
the shell thickness $\xi_{\rm shell}$ (see~\cite{Jinno:2020eqg}) and the efficiency factor $\kappa$ (see~\cite{Giese:2020rtr, Giese:2020znk}) can be determined. 

With this information, the wavenumbers $q_0$, $q_1$, $q_e$ and the momentum-integrated GW power $Q^\prime_{\rm int}$ can be extracted 
from Fig.~\ref{fig:params}. 
$Q^\prime$ can be constructed from (\ref{eq:shape}), with the normalization $S_0$ determined from the condition 
$\int \, d\,{\ln}\,q \, S_f(q) = 1$.
The redshifted energy density in GWs today is then according to (\ref{eq:OmQp}) and~\cite{Caprini:2019egz} given by 
\be
\Omega_{\rm GW} = F_{\rm GW,0} \, \frac{4H \tau_{\rm sw}}{3\pi^2} \frac{H}{\beta} \times Q' \, ,
\ee
with 
\be
F_{\rm GW,0} = (3.57 \pm 0.05) \times 10^{-5} 
\left( \frac{100}{g_*} \right)^{1/3} \, ,
\ee
where $g_*$ is the number of relativistic degrees of freedom right after the phase transition.

The spectral shape can be determined via the shape function in (\ref{eq:shape}) and redshifting the 
frequencies via
\be
f_x = 16.5 \times 10^{-3} \, {\rm mHz} \, \frac{q_x}{\beta}\,  \frac{\beta}{H} \,  
\frac{T}{100 \, {\rm GeV}} \,
\left( \frac{g_*}{100} \right)^{1/6} \, ,
\ee
for $q_0$, $q_1$ and $q_e$, respectively.

\section{The IR tail of the spectrum}
\label{sec:IRtail}

\begin{figure} [h]
	\centering
	\includegraphics[width=.6\textwidth]{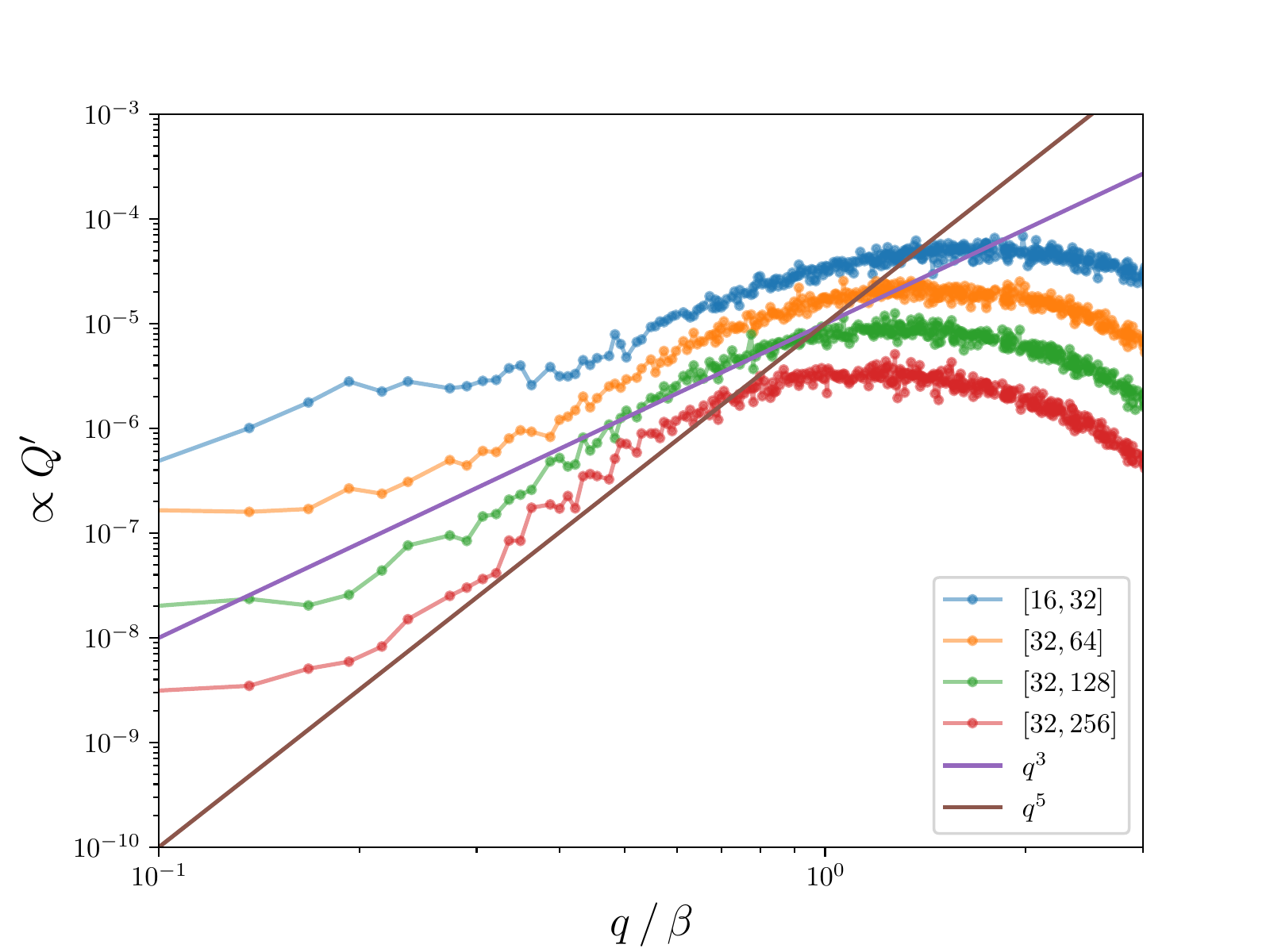} 
	\caption{The IR tail of the GW spectrum for $L=80 v_w/\beta$, intermediate strenght, $v_w=0.8$, $N=512$, and different 
integration times (in units of $1/\beta$). The individual lines are shifted by factors of 2 in order to make them better visible.
	}
	\label{fig:IRtail}
\end{figure}

In this section, we study the IR tail of the spectrum in more detail.
The simulations are run for even larger box size, $L=80 v_w/\beta$ ($\sim$ 20.000 bubbles), intermediate strenght, $v_w=0.8$, and different 
integration times. The results are show in Fig.~\ref{fig:IRtail}.

A straight-forward interpretation of the IR tail of the spectrum is hindered by several issues. 
First, one would like to extract the late-time behavior of the system and start integration late 
enough such that the bubble collisions during percolation give only a minor contribution to the spectrum.
Second, only the modes for which at least $O(10)$ oscillations are tracked in the Fourier transformation with respect to time are correctly captured, 
which is an argument in favor of longer 
integration times. At the same time, when a simulation is run longer than $L/(2c_s)$, the first bubble
will start to interact with its mirror images, leading to IR artifacts. 

Overall, the best compromise for us is to run until $T\sim L$ and to neglect the modes in the deep IR. 
This corresponds to the orange spectrum which seems to scale close to $k^3$ in the IR which motivates
the choice in the main text. Still, even though we deem this part of the spectrum unreliable, one has to note that for longer integration times the spectrum decays a bit steeper in the deep IR. Ultimately, even larger simulations (i.e.~larger box size at fixed grid spacing) are probably required to 
settle this issue. 

\section{$T^{\mu 0}$ conservation test}
\label{sec:T_cons}

\begin{figure} [h]
	\centering
	\includegraphics[width=.6\textwidth]{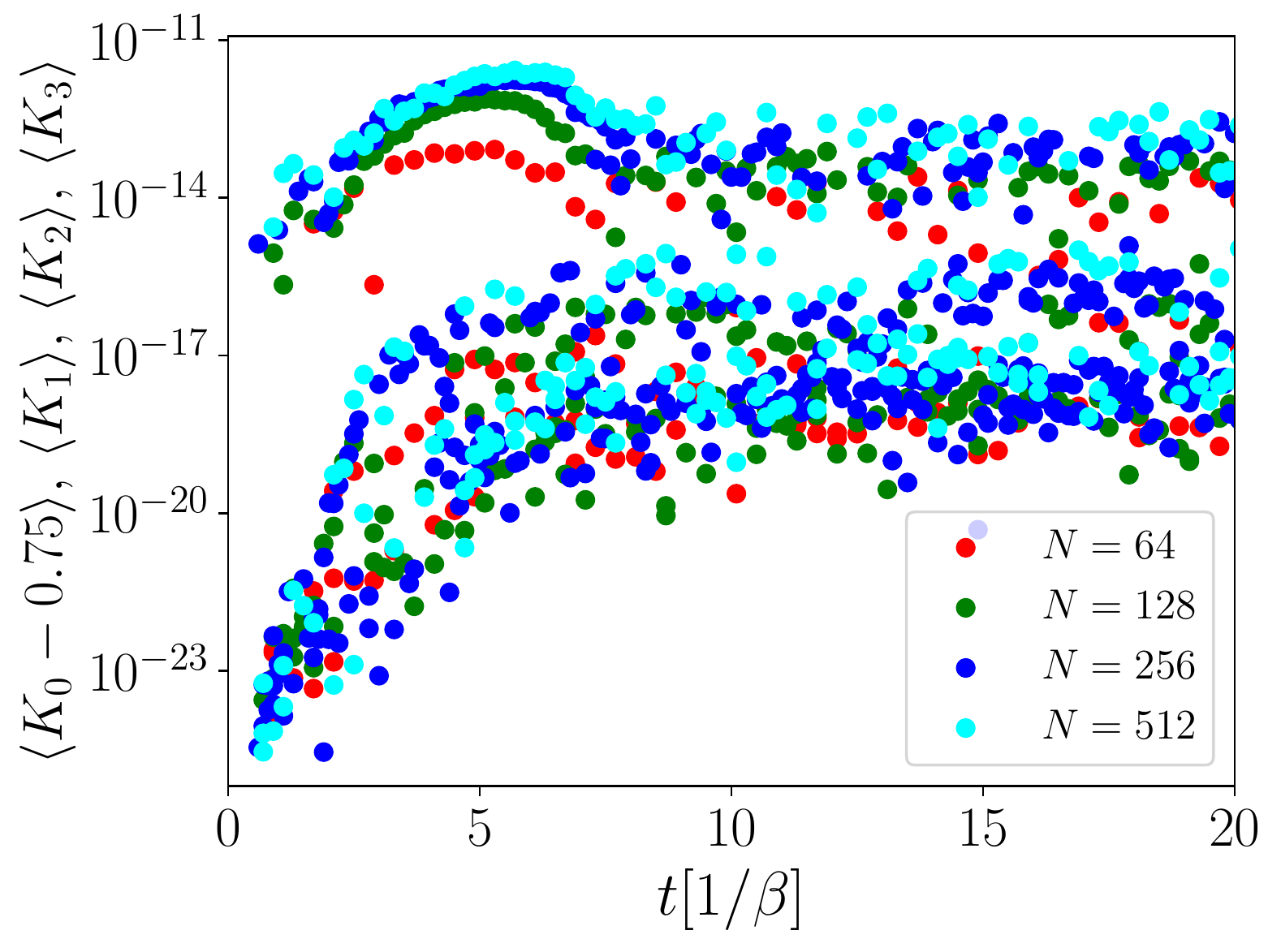} 
	\caption{
		Time evolution of $\langle K^{\mu}\rangle$ for different grid sizes ($N = 64,\, 128,\, 256$ and $512$), $v_w = 0.8$ and intermediate transitions.
	}
	\label{fig:Kmucons}
\end{figure}

In this section, we test the conservation of $K^\mu := T^{\mu0}$. According to the conservation of the stress-energy tensor, Eq.~(\ref{eq:Tcons}), grid average of $K^{\mu}$ must be constant for all times. Also, the numerical scheme described in Sec.~\ref{sec:KTscheme} guarantees the conservation for the fluxes of $K^\mu$ up to the numerical precision of the types used for the fluxes (C++ type {\it double} in our case). 
We display in Fig.~\ref{fig:Kmucons} the average values of each component of $ K^{\mu}$. For $K^0$ we subtract the vacuum contribution.  We see that all of them are conserved up to $10^{-11}$ precision, as expected from the stress energy conservation, Eqs.~\ref{eq:K0cons} and \ref{eq:Kicons}.

\section{Kinetic energy extrapolation}
\label{sec:kinetic}

In this section, we investigate the time evolution of the kinetic energy of the fluid.
Even though the kinetic energy is not conserved per se, it should be conserved as long as fluctuations are small enough that the fluid can be treated as a superposition of 
plane waves. The main purpose here is to assess the impact of the grid spacing on the kinetic energy and study the convergence of the method. 

\begin{figure}
	\centering
	\includegraphics[width=.5\textwidth]{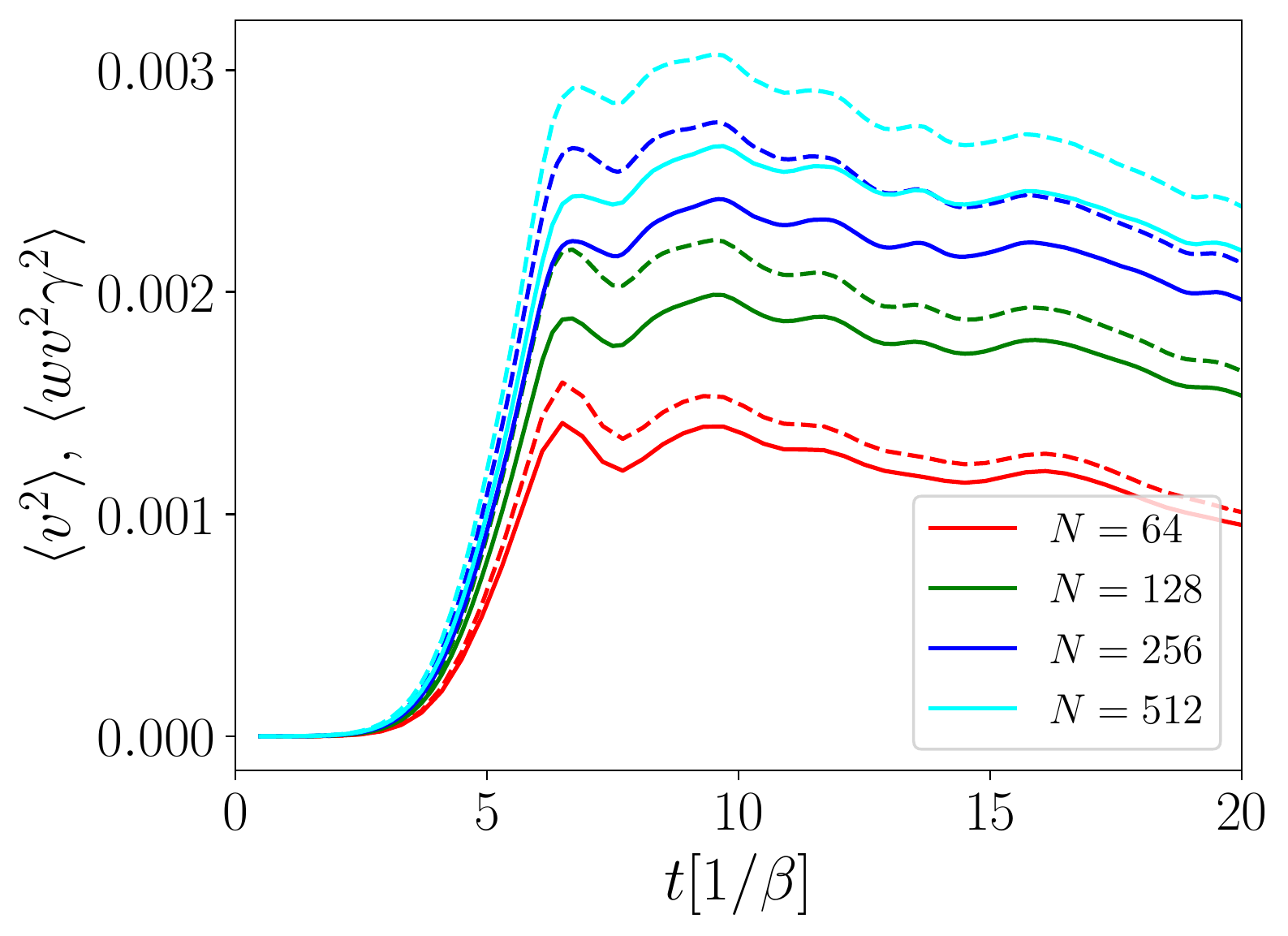} 
	\includegraphics[width=.45\textwidth]{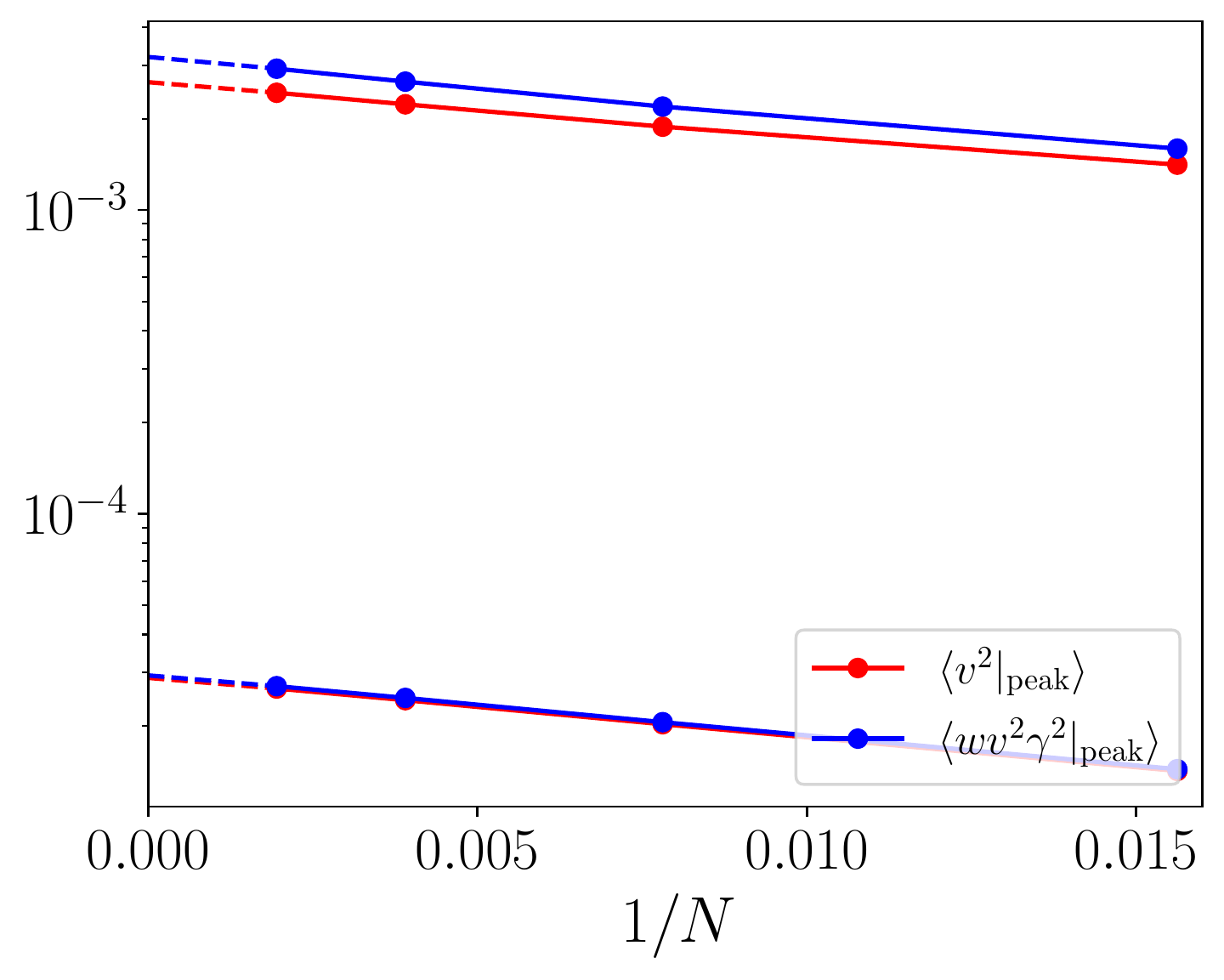} 
	\caption{On the left, we show the time evolution of the kinetic energy for different box size $N$. Those values are for intermediate transitions with $v_w = 0.8$. Dashed lines indicate $\langle wv^2\gamma^2 \rangle$ and solid lines $\langle v^2 \rangle$. On the right, we display the kinetic energy value at the first peak (around $t \simeq 7 /\beta$) as a function of $N$. We include intermediate (top curves) and weak transitions (bottom curves). Dashed lines indicate extrapolation to infinity simulation resolution.}
	\label{fig:kinetic}
\end{figure}

On the left panel of Fig.~\ref{fig:kinetic} we display two probes for the fluid kinetic energy, $v^2$ (solid) and $wv^2\gamma^2$ (dashed) as a function of time for intermediate phase transitions with $v_w=0.8$. Each color indicates a different grid resolution $N$. We notice the saturation of kinetic energy around $t = 7/\beta$, when the phase transition is completed. When increasing the resolution, the kinetic energy also substantially increases, although the difference between $N=256$ and $N=512$ is already small. Since the GW spectrum is roughly proportional to $v^4$, we can extrapolate the kinetic energy to infinite simulation resolution to estimate how much we are underestimating the GW spectrum. On the right panel of Fig.~\ref{fig:kinetic} we display the kinetic energy as a function of the grid resolution for weak (bottom lines) and intermediate transitions (top lines). The dashed lines indicate the extrapolation towards infinity resolution. 
We see that both for weak and intermediate transitions this extrapolation indicates a $10\%$ loss in the kinetic energy, which would mean around $20\%$ underestimation for the GW spectrum.

\section{Full parameter sets}
\label{sec:full}

\begin{figure}
	\centering
	\includegraphics[width=\textwidth]{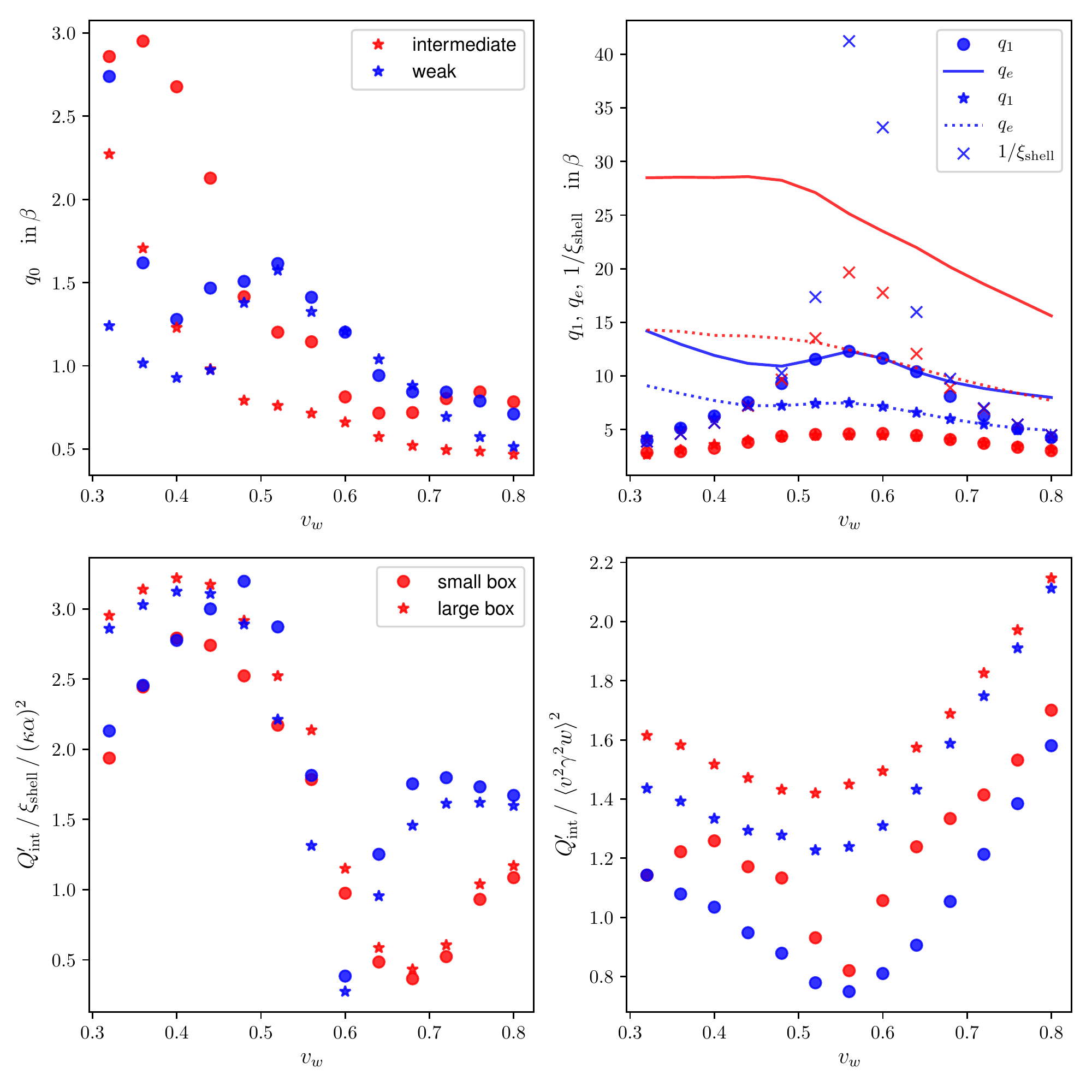} 
	\caption{
		The extracted fitting parameters as functions of the wall velocity. Blue (red) points correspond to phase transitions with weak (intermediate) strength and $\alpha = 0.0046$ ($\alpha = 0.05$). The top left plot shows the scale $q_0$ related to the bubble size, the top right plot the scales related to the shell thickness and the exponential cutoff ($q_1$ and $q_e$), and the bottom plots the amplitude. Dots are from simulations with small box size ($L=20v_w/\beta$) while stars are from simulations 
with large box size ($L=40v_w/\beta$).}
	\label{fig:full_params}
\end{figure}

Figure~\ref{fig:full_params} shows all fit parameters of all simulations we performed. Again, blue points denote weak phase transitions while red points denote phase transitions with intermediate strength. Dots are from simulations with small box size ($L=20v_w/\beta$) while stars are from simulations with large box size ($L=40v_w/\beta$).
The top left panel shows the scale $q_0$ related to the bubble size. Simulations with small box size  significantly 
overestimate $q_0$ and we report only $q_0$ from the simulations with large box size in the main text.
All simulations have been run using $N=512$.

Contrarily, the UV quantities shown in the top right panel are bounded by the exponential cutoff. The proper lines correspond to $q_e$ in the small simulations while the dotted lines show $q_e$ in the simulations with larger box size. While the measurements agree
as long as $q_1 < q_e$, the lower $q_e$ limits the measurement of $q_1$, especially for the weak phase transitions. As discussed in the main text, $q_e$ is lower for weak phase transitions and does not double even if the grid spacing is halved (by halving the box size $L$ and keeping the grid size $N$ fixed). Hence $q_e$ does not only result from the grid spacing.

The bottom panels show the amplitude of the GW spectrum. The amplitude from small box size seems to exceed the one from large box size. Since the power spectrum is reduced due to finite size effects (see Sec.~\ref{sec:res}) and the discrepancies 
are largest for small wall velocities, we chose to report the amplitude from the large box size in the main text. Alternatively, one could choose the bigger of the two in all cases.

\newpage

\bibliographystyle{JHEP}
\bibliography{refs}

\end{document}